\documentclass[9pt, a4paper,twocolumn]{article} 
\usepackage[style=authoryear,backend=biber]{biblatex}
\usepackage{tabularx}
\usepackage{setspace}
\usepackage{subfloat}
\usepackage{mathtools}
\usepackage{amsmath,amsfonts,amssymb,amsthm,mathrsfs, bm}
\usepackage[utf8]{inputenc}
\usepackage[swedish, english]{babel}
\usepackage{csquotes}
\usepackage{subfig}
\usepackage{graphicx}
\usepackage{supertabular}                                             
\usepackage[table]{xcolor} 

\usepackage{graphicx}
\usepackage{epstopdf}
\usepackage{microtype}
\usepackage{multirow}
\usepackage{hyperref}
\usepackage{xcolor}
\usepackage{authblk}
\usepackage{datetime2}	
\usepackage[left=0.5in, right=0.5in, top=0.8in, bottom=0.8in]{geometry} 
\usepackage{setspace}	
\theoremstyle{plain}
\newtheorem{theorem}{Theorem}[section]
\newtheorem{lemma}{Lemma}[section]

\newtheorem{proposition}{Proposition}[section]
\newtheorem{assumption}{Assumption}[section]

\usepackage{amsmath,amsfonts,amssymb,amsthm,mathrsfs, bm}
\usepackage{stmaryrd}

\usepackage{graphicx}
\usepackage{tikz}
\usepackage[utf8]{inputenc}
\usepackage{pgfplots} 
\usepackage{pgfgantt}
\usepackage{pdflscape}
\pgfplotsset{compat=newest} 
\pgfplotsset{plot coordinates/math parser=false}

\usepackage{cancel}
\usepackage{endnotes}
\usepackage{commath}
\usepackage{gensymb}
\usepackage{mathtools}
\usepackage{tikz} 
\usetikzlibrary{calc,quotes,angles}
\usepackage{tkz-euclide}
\usetikzlibrary{automata,positioning} 

\usepackage{xcolor}
\usepackage{tikz,tkz-euclide,siunitx}
\usepackage{etoolbox} 

\usetikzlibrary{quotes,arrows.meta,angles,calc,shadings,positioning,babel}

\newtheorem{remark}{Remark}

\usepackage{endnotes}
\usepackage{commath}
\usepackage{gensymb}

\makeatletter
\patchcmd{\tkz@DrawLine}{\begingroup}{\begingroup\makeatletter}{}{}
\makeatother

\allowdisplaybreaks

\DeclareMathOperator{\RE}{Re}

\DeclareMathOperator{\ran}{ran}
\makeatletter
\newcommand\makebig[2]{%
  \@xp\newcommand\@xp*\csname#1\endcsname{\bBigg@{#2}}%
  \@xp\newcommand\@xp*\csname#1l\endcsname{\@xp\mathopen\csname#1\endcsname}%
  \@xp\newcommand\@xp*\csname#1r\endcsname{\@xp\mathclose\csname#1\endcsname}%
}
\makeatother

\providecommand*{\ped}[1]{%
\ensuremath{_\textnormal{#1}}}

\providecommand*{\eu}%
{\ensuremath{\mathrm{e}}}
\providecommand*{\GammaF}%
{\ensuremath{\mathrm{\Gamma}}}
\providecommand*{\BetaF}%
{\ensuremath{\mathrm{\Beta}}}
\usepackage{bbm}
\makebig{biggg} {3.0}
\makebig{Biggg} {3.5}
\makebig{bigggg}{4.0}
\makebig{Bigggg}{4.5}
\makebig{biggggg}{5.0}
\makebig{Biggggg}{5.5}
\usepackage{longtable}
\usepackage{xcolor}
\DeclareMathSymbol{\Gamma}{\mathalpha}{letters}{"00}
\DeclareMathSymbol{\Delta}{\mathalpha}{letters}{"01}
\DeclareMathSymbol{\Theta}{\mathalpha}{letters}{"02}
\DeclareMathSymbol{\Lambda}{\mathalpha}{letters}{"03}
\DeclareMathSymbol{\Xi}{\mathalpha}{letters}{"04}
\DeclareMathSymbol{\Pi}{\mathalpha}{letters}{"05}
\DeclareMathSymbol{\Sigma}{\mathalpha}{letters}{"06}
\DeclareMathSymbol{\Upsilon}{\mathalpha}{letters}{"07}
\DeclareMathSymbol{\Phi}{\mathalpha}{letters}{"08}
\DeclareMathSymbol{\Psi}{\mathalpha}{letters}{"09}
\DeclareMathSymbol{\Omega}{\mathalpha}{letters}{"0A}

\usepackage{hyperref}
\hypersetup{
    colorlinks = true,
    linkbordercolor = {white},
            linkcolor=blue,
            bookmarks=true,
            filecolor=blue,
            urlcolor=blue,
            citecolor=blue
}
\usepackage{cuted}%
\DeclareMathAlphabet{\mathcal}{OMS}{cmsy}{m}{n}

\newenvironment{automaticaabstract}
{\par\small\noindent\textbf{Abstract.} }
{\par}

\newcommand{\automatickeywords}[1]{%
\par\small\noindent\textit{Keywords:} #1\par
}
\begin{document}
\title{Passivity-exploiting stabilization of semilinear single-track\\ vehicle models with distributed tire friction dynamics}
\date{}
\author[a,b,c]{Luigi Romano\thanks{Corresponding author. Email: luigi.romano@liu.se.}}
\author[b]{Ole Morten Aamo}
\author[c]{Miroslav Krstić}
\author[a]{Jan Aslund}
\author[a]{Erik Frisk}
\affil[a]{\footnotesize{Department of Electrical Engineering, Linköping University, SE-581 83 Linköping, Sweden}}
\affil[b]{\footnotesize{Department of Engineering Cybernetics, Norwegian University of Science and Technology, O. S. Bragstads plass 2, NO-7034, Trondheim, Norway}}
\affil[c]{\footnotesize{Department of Mechanical and Aerospace Engineering, University of California San Diego, La Jolla, CA, 92093, USA}}

\maketitle

\begin{strip}
    \centering
    \begin{minipage}{.8\textwidth}
\begin{automaticaabstract}
          This paper addresses the local stabilization problem for semilinear single-track vehicle models with distributed tire friction dynamics, represented as interconnections of ordinary differential equations (ODEs) and hyperbolic partial differential equations (PDEs). A passivity-exploiting backstepping design is presented, which leverages the strict dissipativity properties of the PDE subsystem to achieve exponential stabilization of the considered ODE-PDE interconnection around a prescribed equilibrium. Sufficient conditions for local well-posedness and exponential convergence are derived by constructing a Lyapunov functional combining the lumped and distributed states. Both state-feedback and output-feedback controllers are synthesized, the latter relying on a cascaded observer. The theoretical results are corroborated with numerical simulations, considering non-ideal scenarios and accounting for external disturbances and uncertainties. Simulation results confirm that the proposed control strategy can effectively and robustly stabilize oversteer vehicles at high speeds, demonstrating the relevance of the approach for improving the safety and performance in automotive applications.
\end{automaticaabstract}
        \hspace{0.5cm}

\automatickeywords{Vehicle dynamics; distributed friction models; distributed parameter systems; hyperbolic ODE-PDE systems; semilinear systems; backstepping control}
    \end{minipage}
\end{strip}


\section{Introduction}
\label{sec:introduction}

The safe and stable operation of road vehicles has long been a central focus in automotive engineering. Loss of stability during high-speed maneuvers, such as sharp cornering, lane changes, or evasive braking, can severely compromise vehicle safety and performance. In particular, lateral instabilities such as oversteer or oscillatory yaw motions are critical risk factors in accident scenarios. For this reason, modern automotive control systems -- ranging from electronic stability control (ESC) to advanced driver-assistance systems (ADAS) -- rely on robust stabilization strategies that ensure predictable and reliable handling characteristics \parencite{LateralControl,Savaresi}.

A key determinant of vehicle stability lies in the interaction between the tire and the road surface. Rolling contact phenomena occurring inside the tire's contact patch govern the transmission of lateral and longitudinal forces, directly influencing the vehicle's ability to accelerate, decelerate, and follow desired paths \parencite{LibroMio,Pacejka2}. Traditional vehicle models, such as the classic \emph{single-track} representation \parencite{Pacejka2,Guiggiani}, approximate these forces using lumped descriptions, in which the tire forces are modeled as static nonlinearities. Whilst such simplifications have been instrumental for controller design and industry adoption \parencite{Gerdes3,IEEEVT1,IEEEVT2,IEEEVT3,LuGreControl2}, they fail to capture the inherently distributed nature of the tire-road interaction. In fact, as extensively discussed in \textcite{Takacs2,Takacs1,Takacs3,Takacs5,Beregi1,Beregi3,BicyclePDE}, both theoretical studies and experimental evidence have demonstrated that the distributed dynamics of the tire is responsible for introducing delays, memory effects, and nonlinear behaviors that cannot be adequately captured by lumped descriptions.

Distributed tire friction models, such as the Dahl and LuGre formulations developed in \textcite{TsiotrasConf,Tsiotras1,Tsiotras2,Deur0,Deur1,Deur2}, address this limitation by representing tire forces as the outcome of spatially distributed dynamics of bristle-like elements. In these models, hyperbolic partial differential equations (PDEs) describe the evolution of internal deflections or state variables across the contact patch, which are then integrated to yield the net contact forces acting on the vehicle. As a result, when such models are coupled with the rigid-body equations for the lateral vehicle motion, the overall dynamics are described by an ODE-PDE interconnection. This structure offers a far more accurate and predictive description of the tire-vehicle behavior, particularly under transient conditions such as sudden steering inputs or road disturbances. However, it also complicates the design of stabilizing controllers, since the resulting system is infinite-dimensional and nonlinear. In previous works, these drawbacks have been partly overcome by replacing the PDE dynamics with lumped approximations, enabling the design of control strategies for traction and braking \parencite{Horowitz1,Horowitz2,Horowitz3,Horowitz4}. However, to the best of the authors' knowledge, with the unique exception of \textcite{MioTITS}, the development of rigorous stabilization algorithms for nonlinear vehicle models with distributed tire friction, similar to those introduced in \textcite{SemilinearV}, remains unexplored.

In this context, the synthesis of control and estimation algorithms considering infinite-dimensional vehicle models becomes an extremely appealing topic for automotive research, especially in light of the recent advancements in PDE control. Indeed, over the last two decades, systematic tools have been developed for the stabilization of hyperbolic PDEs and ODE-PDE interconnections. For systems with bounded control operators, early contributions employed linear-quadratic (LQ) optimal control strategies \parencite{LQ1,LQ2}, also based on the classic results contained in \textcite{Weiss,Zwart}. For problems with unbounded input and/or measurements, the backstepping method has proven to be a versatile and powerful approach, leading to a proliferation of contributions concerning observer and controller design \parencite{Krstic00,Krstic0,Krstic1,Krstic2,Krstic3}, and becoming the dominant methodology for boundary-actuated PDE and ODE-PDE systems. Over the years, PDE backstepping has been successfully extended to observer design, parameter estimation, and the stabilization of systems with time-varying delays or cascaded PDE structures, such as those considered in \textcite{Ole00,Ole0,Bekiaris,Bresch,Auriol}. Lyapunov-based analyses, as performed in \textcite{Ole1,Ole2,Ole3}, have often complemented these designs by providing constructive stability guarantees and conditions for exponential convergence. Robust control methods, allowing for accurate output tracking, have been presented in \textcite{Cristofides} concerning linear and linearized hyperbolic and parabolic PDEs.

Limited to linear single-track models, similar techniques to those illustrated in \textcite{Cristofides} were indeed employed in \textcite{MioTITS} to achieve motion tracking with the desired level of performance. For semilinear systems as those considered in this manuscript, however, these methods are not immediately applicable, due to the complex structure of the nonlinearities that incorporate the input term. Moreover, the strategy presented in \textcite{MioTITS,Cristofides} requires analyzing the closed loop behavior of the PDE subsystem \emph{a posteriori}, without providing direct indications about stability in the desired functional space. 
In semilinear single-track models, these difficulties are exacerbated by the friction nonlinearities, which are often non-smooth, non-Lipschitz, and both state and input-dependent. In fact, the intricate nature of the ODE-PDE interconnection governing the dynamics of semilinear single-track vehicle models seems to require a tailored strategy that leverages the peculiar features of the tire-road rolling contact mechanics.
In this context, the present work develops an \emph{ad-hoc} passivity-exploiting stabilization method for all-wheel-steering vehicles. The central insight of the proposed approach is that the PDE subsystem often possesses strict dissipativity properties, as discussed in \textcite{DistrLuGre,FrBD}. Indeed, being essentially governed by frictional mechanisms, the tire-road rolling contact process inherently dissipates energy in the form of slip losses. By strategically leveraging this property, a backstepping controller is synthesized that ensures local exponential stabilization of the coupled ODE-PDE system around prescribed equilibria. For linear single-track models, global results are derived. From an implementation perspective, compared to the approach proposed in \textcite{MioTITS}, which relies on the injection of boundary terms and is limited to linear models, the controller synthesized in this work uses only bounded functionals of the distributed states, achieving stabilization without altering the semilinear structure of the ODE-PDE interconnection. 
The theoretical development is also corroborated by numerical simulations accounting for realistic operating conditions that incorporate parametric uncertainties and external disturbances.

From an application standpoint, the proposed control framework is directly relevant for enhancing vehicle safety and performance. By explicitly accounting for the distributed nature of tire-road interaction, the developed method provides a principled way to stabilize vehicles subject to complex contact dynamics, particularly under oversteer conditions at high speeds, and in the presence of micro-shimmy oscillations \parencite{Takacs2,Takacs1,Takacs3,Takacs5,Beregi1,Beregi3,BicyclePDE}. This represents a step beyond conventional control strategies that rely on lumped tire models, offering new opportunities for advanced stability augmentation and predictive control in automotive systems.

The remainder of this paper is organized as follows. Section \ref{sect:Problem} formulates the problem and introduces the main structural assumptions. The state and output-feedback stabilization strategies are then presented in Section \ref{sect:Controllllll}. Section \ref{sect:sim} exemplifies the proposed approach considering realistic scenarios accounting for parametric uncertainties and external disturbances. Finally, Section \ref{sect:conclusion} concludes the paper and outlines future research directions.


\subsection*{Notation}
In this paper, $\mathbb{R}$ denotes the set of real numbers; $\mathbb{R}_{>0}$ and $\mathbb{R}_{\geq 0}$ indicate the set of positive real numbers excluding and including zero, respectively. 
The set of $n\times m$ matrices with values in $\mathbb{F}$ ($\mathbb{F} = \mathbb{R}$, $\mathbb{R}_{>0}$, or $\mathbb{R}_{\geq0}$) is denoted by $\mathbf{M}_{n\times m}(\mathbb{F})$ (abbreviated as $\mathbf{M}_{n}(\mathbb{F})$ whenever $m=n$). $\mathbf{GL}_n(\mathbb{F})$ and $\mathbf{Sym}_n(\mathbb{F})$ represents the groups of invertible and symmetric matrices, respectively, with values in $\mathbb{F}$; the identity matrix on $\mathbb{R}^n$ is indicated with $I_n$. A positive-definite matrix is noted as $\mathbf{M}_n(\mathbb{R}) \ni Q \succ 0$.
The standard Euclidean norm on $\mathbb{R}^n$ is indicated with $\norm{\cdot}_2$; matrix norms are simply denoted by $\norm{\cdot}$.
$L^2((0,1);\mathbb{R}^n)$ denotes the Hilbert space of square-integrable functions on $(0,1)$ with values in $\mathbb{R}^n$, endowed with inner product $\langle \zeta_1, \zeta_2 \rangle_{L^2((0,1);\mathbb{R}^n)} = \int_0^1 \zeta_1^{\mathrm{T}}(\xi)\zeta_2(\xi) \dif \xi$ and induced norm $\norm{\zeta(\cdot)}_{L^2((0,1);\mathbb{R}^n)}$. The Hilbert space $H^1((0,1);\mathbb{R}^n)$ consists of functions $\zeta\in L^2((0,1);\mathbb{R}^n)$ whose weak derivative also belongs to $L^2((0,1);\mathbb{R}^n)$; it is naturally equipped with norm $\norm{\zeta(\cdot)}_{H^1((0,1);\mathbb{R}^n)}^2 \triangleq \norm{\zeta(\cdot)}_{L^2((0,1);\mathbb{R}^n)}^2 + \norm{\pd{\zeta(\cdot)}{\xi}}_{L^2((0,1);\mathbb{R}^n)}^2$. For a matrix-valued function $K(\xi)$, $\norm{K(\cdot)}_\infty \triangleq \sup_{\xi \in [0,1]}\norm{K(\xi)}$. $C^k([0,T];\mathcal{Z})$ ($k \in \{1, 2, \dots, \infty\}$) denotes the space of $k$-times continuously differentiable functions on $[0,T]$ with values in $\mathcal{Z}$ (for $T = \infty$, the interval $[0,T]$ is identified with $\mathbb{R}_{\geq 0}$). Given two Hilbert spaces $\mathcal{V}$ and $\mathcal{W}$, $\mathscr{L}(\mathcal{V};\mathcal{W})$ denotes the spaces of linear operators from $\mathcal{V}$ to $\mathcal{W}$ (abbreviated $\mathscr{L}(\mathcal{V})$ if $\mathcal{V} = \mathcal{W}$). Finally, the spectrum of a possibly unbounded operator $(\mathscr{O},\mathscr{D}(\mathscr{O}))$ with domain $\mathscr{D}(\mathscr{O})$ is denoted by $\sigma(\mathscr{O})$.

\section{Model description and preliminaries}\label{sect:Problem}
This section is dedicated to introducing the considered family of semilinear single-track models, along with the main assumptions formulated about their dynamics. In particular, the governing equations of the model are presented in Section \ref{sect:modelSed}, whereas mild structural assumptions are postulated in Section \ref{sect:ass}, where some preliminary results are also collected.

\subsection{Model description}\label{sect:modelSed}
In the following, Section \ref{sect:exampleEqs} reviews the governing equations of the semilinear single-track models to the extent that is necessary to understand the manuscript, whereas Section \ref{sect:stateSpace}  introduces a compact state-space representation more amenable to mathematical analysis. 

\subsubsection{Lateral vehicle dynamics with distributed tire friction dynamics}\label{sect:exampleEqs} 
\begin{figure}
\centering
\includegraphics[width=0.8\linewidth]{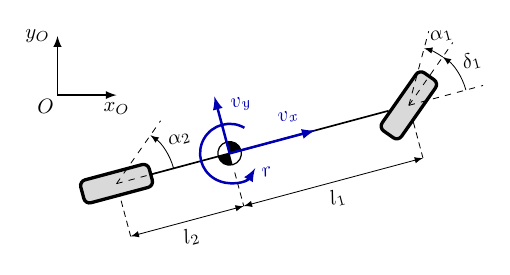} 
\caption{Single-track vehicle model.}
\label{figureForcePostdoc}
\end{figure}

As illustrated schematically in Figure \ref{figureForcePostdoc}, this paper examines semilinear single-track models that govern the lateral dynamics of a road vehicle traveling at a constant cruising speed, and subjected to slow-varying wind disturbances. The model presented here is adapted from \textcite{SemilinearV}.
In particular, for sufficiently small steering inputs, the linear ODE describing the rigid vehicle dynamics may be deduced to be \parencite{Guiggiani}
\begin{subequations}\label{eq:rigid}
\begin{align}
\dot{v}_y(t) & = -\dfrac{1}{m}\bigl(F_{y1}(t) + F_{y2}(t) - F\ped{w}\bigr) -v_xr(t), \\
\dot{r}(t) & = -\dfrac{1}{I_z}\bigl( l_1F_{y1}(t)-l_2F_{y2}(t)-l\ped{w}F\ped{w}\bigr), && t\in (0,T),
\end{align}
\end{subequations}
where the lumped states $v_y(t)$, $r(t) \in \mathbb{R}$ are the vehicle's lateral velocity and yaw rate, $v_x\in \mathbb{R}_{>0}$ is its constant longitudinal speed, $m\in \mathbb{R}_{>0}$ and $I_z\in \mathbb{R}_{>0}$ denote respectively the vehicle mass and moment of inertia of the center of gravity around the vertical axis, and $l_1$, $l_2 \in \mathbb{R}_{>0}$ are the front and rear axle lengths. The external force $F\ped{w} \in \mathbb{R}$ represents a constant or slow-varying perturbation term generated by a lateral wind gust, and $l\ped{w} \in \mathbb{R}$ denotes the offset of its point of application from the center of gravity \parencite{Guiggiani}. In turn, adopting a distributed model for dry or lubricated friction \parencite{DistrLuGre}, the tire forces $F_{y1}(t)$, $F_{y2}(t) \in \mathbb{R}$ may be calculated as
\begin{align}\label{eq:Fi}
\begin{split}
F_{yi}(t) & = F_{zi}\int_0^1 \bar{p}_i(\xi)\sigma_{i}z_i(\xi,t)  \dif \xi, \quad i \in \{1,2\},
\end{split} 
\end{align}
where the distributed state $z_i(\xi,t) \in \mathbb{R}$, $i = \{1,2\}$, represents the deflection of a bristle element schematizing a tire rubber particle or an asperity inside the contact patch, $F_{zi} \in \mathbb{R}_{>0}$ denotes the vertical force acting on the tire, $\bar{p}_i \in C^1([0,1];\mathbb{R}_{\geq 0})$ is the nondimensional vertical pressure distribution, and $\sigma_{i} \in \mathbb{R}_{>0}$ is the normalized micro-stiffness coefficient \parencite{DistrLuGre}. The bristle dynamics obeys the following semilinear PDE \parencite{SemilinearV}:  \begin{small}
\begin{subequations}\label{Eq:PDEz}
\begin{align}\label{Eq:PDEzDKKF}
\begin{split}
& \dpd{z_i(\xi,t)}{t} + \dfrac{v_x}{L_i}\dpd{z_i(\xi,t)}{\xi} =2 \phi_i v_i\bigl(v_y(t),r(t), \delta_i(t)\bigr)\\
& \quad -\theta\dfrac{\sigma_{i}\abs{v_i\bigl(v_y(t),r(t), \delta_i(t)\bigr)}_\varepsilon}{\mu_i\bigl(v_i\bigl(v_y(t),r(t), \delta_i(t)\bigr)\bigr)}\Biggl(z_i(\xi,t)-\psi_i\int_0^1\bar{p}_i(\xi)z_i(\xi,t) \dif \xi\Biggr)  \\
& \quad + v_x\dfrac{\psi_i}{L_i}\Biggl( \bar{p}_i(1)z_i(1,t) - \int_0^1 \dod{\bar{p}_i(\xi)}{\xi}z_i(\xi,t)\dif \xi\Biggr), \\
& \qquad \qquad \qquad \qquad \qquad \qquad (\xi,t) \in (0,1)\times (0,T),
\end{split} \\
& z_i(0,t) = 0, \label{eq:BCz}
\end{align}
\end{subequations} \end{small} 
where $L_i \in \mathbb{R}_{>0}$ denotes the contact patch length, $\bar{\mu}_i \in C^0(\mathbb{R};[\mu\ped{min},\infty))$, with $\mu\ped{min}\in \mathbb{R}_{>0}$, the friction coefficient, $\theta \in \mathbb{R}_{\geq 0}$ is a parameter that accounts for variations in friction, the constants $\phi_i \in (0,1]$ and $\psi_i \in [0,1)$ are structural parameters connected with the flexibility of the tire carcass, identically satisfying $\phi_i + \psi_i = 1$, $i \in \{1,2\}$, and the function $\abs{\cdot}_\varepsilon \in C^0(\mathbb{R};\mathbb{R}_{\geq 0})$ denotes the (possibly regularized\footnote{It is common in engineering practice to replace the absolute value with differentiable functions $\abs{\cdot}_\varepsilon \in C^1(\mathbb{R};\mathbb{R}_{\geq 0})$ \parencite{Rill,Rill0}, such as $\abs{v}_\varepsilon \triangleq \sqrt{v^2 + \varepsilon}$, for some $\varepsilon \in \mathbb{R}_{>0}$. This paper considers $\varepsilon \in \mathbb{R}_{\geq 0}$.}) absolute value, converging uniformly to $\abs{\cdot}$ in $C^0(\mathbb{R};\mathbb{R}_{\geq 0})$ for $\varepsilon \to 0$. Distributed friction models accommodated by \eqref{eq:Fi} and \eqref{Eq:PDEz} include the Dahl model, as well as the LuGre and FrBD models in the absence of internal damping\footnote{Incorporating internal damping terms would lead to a more complex structure of the resulting ODE-PDE system. In this context, it is worth mentioning that damping effects were found to be minor, if not completely negligible, in the literature \parencite{TsiotrasConf,Tsiotras1,Tsiotras2,Deur0,Deur1,Deur2}. Extensions to more sophisticated formulations accounting for internal damping might, however, be explored in future works.}.

The rigid relative velocities in \eqref{Eq:PDEzDKKF} are given by
\begin{subequations}\label{eq:slipAngles}
\begin{align}
\begin{split}
v_1\bigl(v_y(t),r(t), \delta_1(t)\bigr)  & = v_x \alpha_1\bigl(v_y(t),r(t), \delta_1(t)\bigr)\\
& = v_y(t) + l_1r(t)-v_x\delta_1(t), 
\end{split} \\
\begin{split}
v_2\bigl(v_y(t),r(t), \delta_2(t)\bigr) &= v_x\alpha_2\bigl(v_y(t),r(t), \delta_2(t)\bigr)\\
& = v_y(t)-l_2r(t)-v_x\delta_2(t),
\end{split}
\end{align}
\end{subequations}
being $\delta_1(t), \delta_2(t) \in \mathbb{R}$ the steering inputs at the front and rear axles, and $\alpha_1(v_y(t),r(t), \delta_1(t)), \alpha_2(v_y(t),r(t), \delta_2(t)) \in \mathbb{R}$ the apparent slip angles, respectively.

Finally, the nondimensional vertical pressure distribution inside the tires' contact patches, appearing in \eqref{eq:Fi} and \eqref{eq:MatricesOOO}, may be modeled using exponentially decreasing functions of the type \parencite{DistrLuGre}
\begin{align}\label{eq:PressureDistr}
\bar{p}_i(\xi) = \bar{p}_{0,i}\exp(-a_i\xi), \quad \xi \in [0,1],
\end{align}
with $a_i \in \mathbb{R}_{>0}$, $i \in \{1,2\}$, and
\begin{align}
\bar{p}_{0,i} & \triangleq \dfrac{a_i}{1-\eu^{-a_i}}, \quad i \in \{1,2\}.
\end{align}
Specifying the contact pressure as in \eqref{eq:PressureDistr} is important but not essential for the scope of this paper. Indeed, other choices of functions $\bar{p}_i(\cdot)$, $i \in \{1,2\}$, are viable, provided that the assumptions of Section \ref{sect:ass} are satisfied. Ultimately, the use of exponentially decreasing profiles as in \eqref{eq:PressureDistr} is also legitimized by its adoption in previous works \parencite{TsiotrasConf,Tsiotras1,Tsiotras2,Deur0,Deur1,Deur2}. More generally, employing decreasing distributions along the contact patch length may be justified by observing that, in automotive tires, centrifugal and viscoelastic effects tend to shift the maximum normal pressure toward the leading edge \parencite{Nikravesh}. 

Equations \eqref{eq:rigid}-\eqref{eq:PressureDistr} completely determine the lateral motion of the single-track model. As explained next in Section \ref{sect:stateSpace}, they may be restated in a compact form which is more amenable to mathematical analysis.

\subsubsection{State-space representation}\label{sect:stateSpace} 
Defining $\mathbb{R}^2 \ni X(t) \triangleq [v_y(t)\; r(t)]^{\mathrm{T}}$, $\mathbb{R}^2\ni z(\xi,t) \triangleq [z_1(\xi,t) \; z_2(\xi,t)]^{\mathrm{T}}$, $\mathbb{R}^2 \ni U(t) \triangleq [\delta_1(t)\; \delta_2(t)]^{\mathrm{T}}$, $\mathbb{R}^2 \ni v(X(t),U(t)) = [v_1(X(t),U(t))\; v_2(X(t),U(t))]^{\mathrm{T}} \triangleq [v_1(v_y(t),r(t), \delta_1(t)) \; v_2(v_y(t),r(t), \delta_2(t))]^{\mathrm{T}}$, and $\mathbb{R}^2 \ni b \triangleq [\frac{F\ped{w}}{m} \; \frac{l\ped{w}F\ped{w}}{I_z}]^{\mathrm{T}}$, \eqref{eq:rigid}-\eqref{eq:PressureDistr} may be recast in the form\footnote{Alternatively, the term $b$ in \eqref{eq:originalSystemsODE} may also model disturbances generated by road banking, for instance by specifying $b = [g\sin\vartheta\cos\phi\; 0]^{\mathrm{T}}$, where $\vartheta$ denotes the bank angle, and $\phi$ the angle between the heading of the vehicle and the tangent to the road path \parencite{Guiggiani2}.}
\begin{subequations}\label{eq:originalSystems}
\begin{align}
\begin{split}
& \dot{X}(t) =A_1X(t)+ G_1(\mathscr{K}_1z)(t)+b, \quad t \in (0,T),
\end{split} \label{eq:originalSystemsODE}\\
\begin{split}
& \dpd{z(\xi,t)}{t} + \Lambda \dpd{z(\xi,t)}{\xi} = \theta\Sigma\Bigl(v\bigl(X(t),U(t)\bigr)\Bigr) \\
& \qquad \qquad \times  \bigl[z(\xi,t) + (\mathscr{K}_2z)(t)\bigr]  +(\mathscr{K}_3z)(t) \\
& \qquad \qquad  +Hv\bigl(X(t),U(t)\bigr),\quad (\xi,t) \in (0,1) \times (0,T),
\end{split} \label{eq:originalSystemsPDE} \\
& z(0,t) = 0, \quad t \in (0,T),\label{eq:originalSystemsBC}
\end{align}
\end{subequations}
where the rigid relative velocity $v \in C^1(\mathbb{R}^{4};\mathbb{R}^{2})$ may be expressed as
\begin{align}\label{eq:relVel}
v(X,U) \triangleq A_2X + G_2U.
\end{align}
In \eqref{eq:originalSystems} and \eqref{eq:relVel}, the matrix $\mathbf{GL}_{2}(\mathbb{R})\cap \mathbf{Sym}_{2}(\mathbb{R})\ni \Lambda \succ 0$ collects the transport velocities, $\Sigma \in C^0(\mathbb{R}^{2};\mathbf{M}_{2}(\mathbb{R}))$ represents the nonlinear source matrix, $A_1, A_2 \in  \mathbf{M}_{2}(\mathbb{R})$, and $G_1, G_2 \in \mathbf{GL}_{2}(\mathbb{R})$, $H \in \mathbf{GL}_{2}(\mathbb{R})\cap \mathbf{Sym}_{2}(\mathbb{R})$ are matrices with constant coefficients, and the operators $(\mathscr{K}_1\zeta)$, $(\mathscr{K}_2\zeta)$, and $(\mathscr{K}_3\zeta)$ satisfy $\mathscr{K}_1, \mathscr{K}_2 \in \mathscr{L}(L^2((0,1);\mathbb{R}^{2});\mathbb{R}^{2})$, and $\mathscr{K}_3\in \mathscr{L}(H^1((0,1);\mathbb{R}^{2});\mathbb{R}^{2})$, with
\begin{subequations}\label{eq:operatorK}
\begin{align}
(\mathscr{K}_1\zeta) &\triangleq   \int_0^1 K_1(\xi) \zeta(\xi) \dif \xi, \\
(\mathscr{K}_2\zeta) &\triangleq   \int_0^1 K_2(\xi)  \zeta(\xi) \dif \xi, \\
(\mathscr{K}_3\zeta) &\triangleq \int_0^1 K_3(\xi)  \zeta(\xi) \dif \xi+K_4\zeta(1),
\end{align}
\end{subequations}
where $K_1\in  C^0([0,1];\mathbf{GL}_{2}(\mathbb{R})) \cap C^0([0,1];\mathbf{Sym}_{2}(\mathbb{R}))$, $K_2, K_3\in C^0([0,1];\mathbf{M}_{2}(\mathbb{R}))$, and $K_4 \in \mathbf{M}_{2}(\mathbb{R})$, and \begin{small}
\begin{align}\label{eq:MatricesOOO}
A_1 & \triangleq \begin{bmatrix} 0 & -v_x \\ 0 &  0 \end{bmatrix}, \quad A_2 \triangleq \begin{bmatrix}1 & l_1 \\ 1 & -l_2 \end{bmatrix},  \quad
G_1 \triangleq -\begin{bmatrix}\dfrac{1}{m} & \dfrac{1}{m} \\ \dfrac{l_1}{I_z} & -\dfrac{l_2}{I_z} \end{bmatrix}, \nonumber\\
 G_2 & \triangleq -v_xI_2, \quad K_1(\xi) \triangleq \begin{bmatrix} F_{z1}\sigma_{1}\bar{p}_1(\xi) & 0 \\ 0 & F_{z2}\sigma_{2}\bar{p}_2(\xi)\end{bmatrix}, \nonumber\\
 K_2(\xi) & \triangleq -\begin{bmatrix} \psi_{1}\bar{p}_1(\xi) & 0 \\ 0 & \psi_{2}\bar{p}_2(\xi)\end{bmatrix}, \nonumber\\
 K_3(\xi) & \triangleq -v_x\begin{bmatrix} \dfrac{\psi_1}{L_1}\dod{\bar{p}_1(\xi)}{\xi} & 0 \\ 0 & \dfrac{\psi_2}{L_2}\dod{\bar{p}_2(\xi)}{\xi}\end{bmatrix},  \nonumber\\
 K_4(\xi) & \triangleq v_x\begin{bmatrix} \dfrac{\psi_1}{L_1}\bar{p}_1(1) & 0 \\ 0 & \dfrac{\psi_2}{L_2}\bar{p}_2(1)\end{bmatrix},  \quad \Lambda \triangleq \begin{bmatrix} \dfrac{v_x}{L_1} & 0 \\ 0 & \dfrac{v_x}{L_2}\end{bmatrix}, \nonumber\\
\Sigma(v) & \triangleq \begin{bmatrix} -\dfrac{\sigma_{1}\abs{v_1}_\varepsilon}{\mu_1(v_1)} & 0 \\0 & -\dfrac{\sigma_{2}\abs{v_2}_\varepsilon}{\mu_2(v_2)} \end{bmatrix}, \quad
H \triangleq 2\begin{bmatrix}\phi_1 & 0 \\ 0 & \phi_2\end{bmatrix}.
\end{align}\end{small}

Equations \eqref{eq:originalSystems}-\eqref{eq:MatricesOOO} describe a semilinear hyperbolic ODE-PDE system, where a bounded functional of the distributed states $z(\xi,t)$ acts as the input to the ODE subsystem.
From a mathematical perspective, the ODE-PDE interconnection \eqref{eq:originalSystems}-\eqref{eq:MatricesOOO} is (locally) well-posed. In particular, this paper considers the Hilbert space $\mathcal{X}\triangleq \mathbb{R}^{2}\times L^2((0,1);\mathbb{R}^{2})$, equipped with norm $\norm{(Z, \zeta(\cdot))}_{\mathcal{X}}^2 \triangleq \norm{Z}_{2}^2 + \norm{\zeta(\cdot)}_{L^2((0,1);\mathbb{R}^{2})}^2$. Theorem \ref{thm:mild} below enounces local well-posedness results for the \emph{mild solution} of \eqref{eq:originalSystems}.

\begin{figure}
\centering
\includegraphics[width=0.9\linewidth]{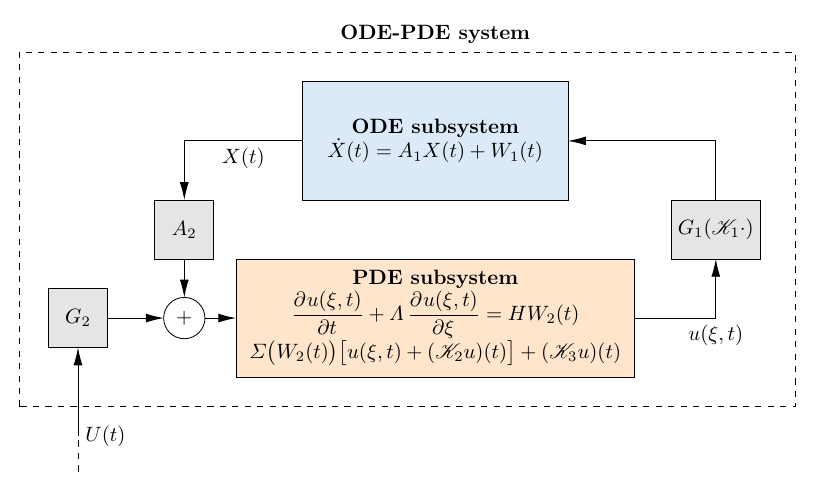} 
\caption{Schematic representation of the ODE-PDE interconnection \eqref{eq:originalSystems}.}
\label{figure:system0}
\end{figure}

\begin{theorem}[Local existence and uniqueness of mild solutions]\label{thm:mild}
Suppose that $\Sigma \in C^0(\mathbb{R}^{2};\mathbf{M}_{2}(\mathbb{R}))$ is locally Lipschitz continuous, and $U \in C^0([0,T];\mathbb{R}^{2})$. Then, for all initial conditions (ICs) $(X_0,z_0) \triangleq (X(0),z(\cdot,0)) \in \mathcal{X}$, there exists $ t\ped{max} \leq \infty$ such that the ODE-PDE system \eqref{eq:originalSystems}-\eqref{eq:MatricesOOO} admits a unique \emph{mild solution} $(X,z) \in C^0([0,t\ped{max});\mathcal{X})$. Moreover, if $t\ped{max} < \infty$, $\norm{(X(t),z(\cdot,t))}_{\mathcal{X}}\to \infty$ for $t \to t\ped{max}$.
\begin{proof}
See the proof of Theorem 3.1 in \textcite{SemilinearV} (for a general statement about semilinear problems, see also Theorems 11.1.5 in \textcite{Zwart} and 6.1.4 in \textcite{Pazy}).
\end{proof}
\end{theorem}
A schematic of the ODE-PDE system \eqref{eq:originalSystems} is illustrated in Figure \ref{figure:system0}, where, for convenience of notation, $\mathbb{R}^{2} \ni W_1(t) \triangleq G_1(\mathscr{K}_1z)(t)+b$ and $\mathbb{R}^{2} \ni W_2(t) \triangleq v(X(t),U(t))$.

As better clarified in Section \ref{sect:ass}, the objective of this paper consists of (locally) stabilizing the hyperbolic ODE-PDE system \eqref{eq:originalSystems} around a target equilibrium $(X^\star, z^\star) \in \mathcal{X}$, starting from available measurements of the rigid relative velocity $v(X(t),U(t))$. In particular, possibly after algebraic manipulations, the measurement output $Y(t) \in \mathbb{R}^{2}$ is supposed to be of the form
\begin{align}\label{eq:measYYYYYY}
Y(t) = Hv(X(t),U(t)).
\end{align}
Measurements of the rigid relative velocity, as described by \eqref{eq:measYYYYYY}, may be acquired either by installing an intelligent tire sensor on each axle, as explained in \textcite{AutoObserv}, or by combining commonly available yaw rate signals with a single accelerometer mounted on the front axle \parencite{MioTITS}. 

\subsection{Assumptions and preliminaries}\label{sect:ass}
To facilitate the observer synthesis, two mild structural assumptions are formulated in this paper. Both relate to the dissipative properties of the friction phenomena that govern the tire-road rolling contact process. In this context, it is essential to introduce the matrix $\mathbf{M}_{2}(\mathbb{R}) \ni \mathscr{Q}(\xi) \triangleq K_1(\xi)H^{-1} $. For every $\xi \in [0,1]$, $\mathscr{Q}(\xi)$ is a symmetric (actually, diagonal) and positive definite matrix, that is, $\mathscr{Q} \in C^0([0,1]; \mathbf{Sym}_{2}(\mathbb{R}))$, and $\mathscr{Q}(\xi) \succ 0$. This implies that $\mathscr{Q}$ defines a positive operator $\mathscr{Q} \in \mathscr{L}(L^2((0,1);\mathbb{R}^{2}))$ on $L^2((0,1);\mathbb{R}^{2})$. Accordingly, Assumption \ref{ass:ODEf} is enounced below.
\begin{assumption}[Strict dissipativity]\label{ass:ODEf} The unbounded operator $(\mathscr{A},\mathscr{D}(\mathscr{A}))$, defined by
\begin{subequations}\label{eq:operatorA}
\begin{align}
(\mathscr{A}\zeta)(\xi) & \triangleq -\Lambda\dpd{\zeta(\xi)}{\xi} + (\mathscr{K}_3\zeta)(\xi), \\
\mathscr{D}(\mathscr{A}) & \triangleq \Bigl\{ \zeta \in H^1((0,1);\mathbb{R}^{2}) \mathrel{\Big|} \zeta(0) = 0\Bigr\}, 
\end{align}
\end{subequations}
satisfies
\begin{align}\label{eq:dissA000}
\begin{split}
& \RE\langle \mathscr{A}\zeta,\mathscr{Q}\zeta\rangle_{L^2((0,1);\mathbb{R}^{2})} \leq -\omega \norm{\zeta(\cdot)}_{L^2((0,1);\mathbb{R}^{2})}^2,
\end{split}
\end{align}
for all $\zeta \in \mathscr{D}(\mathscr{A})$ and some $\omega \in \mathbb{R}_{>0}$.
\end{assumption}
Assumption \ref{ass:ODEf} is satisfied for sufficiently small $\psi_i$, $i \in \{1,2\}$, in \eqref{eq:MatricesOOO} and \eqref{Eq:PDEz}, and obviously for $\psi_i = 0$, as commonly found in the literature \parencite{TsiotrasConf,Tsiotras1,Tsiotras2,Deur0,Deur1,Deur2}. 
It also implies that $(\mathscr{A},\mathscr{D}(\mathscr{A}))$ generates an exponentially stable $C_0$-semigroup on $L^2((0,1);\mathbb{R}^{2})$, ensuring the existence of an inverse operator $\mathscr{A}^{-1}$ on $L^2((0,1);\mathbb{R}^{2})$. Whilst, in general, other choices of positive operators are possible with the given structure of $(\mathscr{A},\mathscr{D}(\mathscr{A}))$, the control strategy developed in the present paper exploits the strict dissipativity properties of the PDE subsystem \eqref{eq:originalSystemsPDE}, and thus requires using explicitly the matrix $\mathscr{Q}(\xi)$.
Before moving to the next assumption, an important remark is formalized below.
\begin{remark}
Typically, for $y \in \mathbb{R}^{2}$, $\ran(\Sigma(\cdot)y + H\cdot) \not = \mathbb{R}^{2}$, which renders it impossible to compensate for all the functionals appearing in \eqref{eq:originalSystemsPDE}. In this context, the operator $(\mathscr{A},\mathscr{D}(\mathscr{A}))$ in \eqref{eq:operatorA} often describes the frictional part of the rolling contact process, which is inherently stable. Therefore, the term $(\mathscr{K}_3z)$ in \eqref{eq:originalSystemsPDE} will not be compensated for by the control action designed in Section \ref{sect:Controllllll}. It is also worth remarking that the operator $(\mathscr{K}_3, H^1((0,1);\mathbb{R}^{2}))$ is generally unbounded on $L^2((0,1);\mathbb{R}^{2})$. Therefore, its direct suppression would make the resulting closed loop ODE-PDE system quasilinear rather than semilinear, due to the presence of the control input $U(t)$ in the matrix $\Sigma(v)$.
\end{remark}

\begin{assumption}[Dissipativity and Lipschitz continuity]\label{ass:ODEf2}
For every $(y,\zeta) \in \mathbb{R}^{2 }\times L^2((0,1);\mathbb{R}^{2 })$, the matrix $\Sigma(y) \in \mathbf{M}_{2}(\mathbb{R})$ satisfies
\begin{align}\label{eq:ineqP}
& \int_0^1 \zeta^{\mathrm{T}}(\xi) \mathscr{Q}(\xi)\Sigma(y)\bigl[\zeta(\xi) + (\mathscr{K}_2\zeta)\bigr] \dif \xi \leq 0.
\end{align}
Additionally, it is globally Lipschitz continuous, that is, there exists $L_{\Sigma}\in \mathbb{R}_{\geq 0}$ such that
\begin{align}\label{eq:SigmaLSigma}
\norm{\Sigma(y_1)-\Sigma(y_2)} & \leq L_{\Sigma}\norm{y_1-y_2}_2,
\end{align}
for all $y_1,y_2 \in \mathbb{R}^{2}$. 
\end{assumption}
Again, Assumption \ref{ass:ODEf2} is fulfilled for sufficiently small $\psi_i$, $i \in \{1,2\}$, in \eqref{eq:MatricesOOO} and \eqref{Eq:PDEz}, and always for $\psi_i = 0$. 
\begin{remark}
Not coincidentally, the dissipativity inequality \eqref{eq:ineqP} ensures global existence and uniqueness for the mild solutions of the open-loop ODE-PDE system \eqref{eq:originalSystems} (see \textcite{SemilinearV}). The global Lipschitz condition \eqref{eq:SigmaLSigma}, which is always verified in practical applications, is instead enforced to facilitate the implementation of an output-feedback controller.
\end{remark}

Together, Assumptions \ref{ass:ODEf} and \ref{ass:ODEf2} permit recovering some preliminary results, as formalized in Propositions \ref{prop2:strictDDD} and \ref{prop:pass}.

\begin{proposition}[Strict dissipativity]\label{prop2:strictDDD}
Suppose that Assumptions \ref{ass:ODEf} and \ref{ass:ODEf2} hold. Then, for all $y \in \mathbb{R}^{2}$, the unbounded operator $(\mathscr{A}_{\Sigma}(y),\mathscr{D}(\mathscr{A}_{\Sigma}(y)))$, defined by
\begin{subequations}
\begin{align}
(\mathscr{A}_{\Sigma}(y)\zeta)(\xi) & \triangleq  \theta\Sigma(y)\bigl[\zeta(\xi) + (\mathscr{K}_2\zeta)\bigr] +(\mathscr{A}\zeta)(\xi), \\
\mathscr{D}(\mathscr{A}_{\Sigma}(y)) & = \mathscr{D}(\mathscr{A}) \triangleq \Bigl\{ \zeta \in H^1((0,1);\mathbb{R}^{2}) \mathrel{\Big|} \zeta(0) = 0\Bigr\}, 
\end{align}
\end{subequations}
also generates an exponentially stable $C_0$-semigroup on $L^2((0,1);\mathbb{R}^{2})$. In particular,
\begin{align}\label{eq:dissA0001}
\begin{split}
& \RE\langle \mathscr{A}_{\Sigma}(y)\zeta,\mathscr{Q}\zeta\rangle_{L^2((0,1);\mathbb{R}^{2})}  \leq -\omega \norm{\zeta(\cdot)}_{L^2((0,1);\mathbb{R}^{2})}^2,
\end{split}
\end{align} 
for all $(y,\zeta) \in \mathbb{R}^{2} \times \mathscr{D}(\mathscr{A}_{\Sigma}(y))$, with the same $\omega \in \mathbb{R}_{>0}$ as in \eqref{eq:dissA000}. 
\begin{proof}
Since $\theta \in \mathbb{R}_{\geq 0}$, the assertion is an immediate consequence of the inequalities \eqref{eq:dissA000} and \eqref{eq:ineqP}.
\end{proof}
\end{proposition}

Assumptions \ref{ass:ODEf} and \ref{ass:ODEf2} combined imply strict passivity for the PDE subsystem \eqref{eq:originalSystemsPDE}-\eqref{eq:originalSystemsBC} with output $(\mathscr{K}_1z)(t)$. It is exactly this intrinsic property that will be exploited in Section \ref{sect:Controllllll} to design a stabilizing control law.
\begin{proposition}[Passivity]\label{prop:pass}
Consider the semilinear PDE 
\begin{subequations}\label{eq:pDpassibity}
\begin{align}
\begin{split}
& \dpd{z(\xi,t)}{t} + \Lambda \dpd{z(\xi,t)}{\xi} = \theta\Sigma\bigl(v(t)\bigr) \bigl[z(\xi,t) + (\mathscr{K}_2z)(t)\bigr] \\
& \qquad \qquad \qquad \qquad \qquad   +(\mathscr{K}_3z)(t)+Hv(t), \\
& \qquad \qquad  \qquad \qquad \qquad  (\xi,t) \in (0,1) \times (0,T),
\end{split} \label{eq:originalSystemsPDEiso} \\
& z(0,t) = 0, \quad t \in (0,T),\label{eq:originalSystemsBCiso}
\end{align}
\end{subequations}
with input $v(t) \in \mathbb{R}^{2}$, and output
\begin{align}\label{eq:outputF}
F(t) \triangleq (\mathscr{K}_1z)(t).
\end{align}
Then, if Assumptions \ref{ass:ODEf} and \ref{ass:ODEf2} hold, the system \eqref{eq:pDpassibity}-\eqref{eq:outputF} is \emph{strictly passive}, and satisfies
\begin{align}\label{eq:passivity}
\begin{split}
& \int_0^t F^{\mathrm{T}}(t^\prime)v(t^\prime) \dif t^\prime \geq V\bigl(z(\cdot,t)\bigr)-V\bigl(z_0(\cdot)\bigr) \\
& \qquad +\omega\int_0^t \norm{z(\cdot,t^\prime)}_{L^2((0,1);\mathbb{R}^{2})}^2 \dif t^\prime, \quad t \in [0,T],
\end{split}
\end{align}
with \emph{storage function}
\begin{align}\label{eq:storage}
V\bigl(z(\cdot,t)\bigr) \triangleq \dfrac{1}{2}\int_0^1 z^{\mathrm{T}}(\xi,t)\mathscr{Q}(\xi)z(\xi,t)\dif \xi,
\end{align}
for all ICs $z_0 \triangleq z(\cdot,0) \in L^2((0,1);\mathbb{R}^{2})$ and inputs $v \in C^0([0,T];\mathbb{R}^{2})$.
\begin{proof}
For simplicity, the result is proved concerning classical solutions $z\in C^1([0,T];L^2((0,1);\mathbb{R}^{2})) \cap C^0([0,T];\mathscr{D}(\mathscr{A}))$, which requires $z_0 \in \mathscr{D}(\mathscr{A})$, $v \in C^1([0,T];\mathbb{R}^{2})$, and possibly $\Sigma \in C^1(\mathbb{R}^{2};\mathbf{M}_{2}(\mathbb{R}))$.
By definition,
\begin{align}\label{eq:Fout0}
\begin{split}
F^{\mathrm{T}}(t)v(t) & = v^{\mathrm{T}}(t)(\mathscr{K}_1z)(t) = \int_0^1 v^{\mathrm{T}}(t)H\mathscr{Q}(\xi)z(\xi,t)\dif \xi \\
& = -\RE\langle\mathscr{A}_{\Sigma}(v(t))z(\cdot,t),\mathscr{Q}z(\cdot,t)\rangle_{L^2((0,1);\mathbb{R}^{2})} \\
& \quad + \dfrac{1}{2}\dod{}{t}\int_0^1 z^{\mathrm{T}}(\xi,t)\mathscr{Q}(\xi)z(\xi,t)\dif \xi.
\end{split}
\end{align}
Consequently, invoking Assumptions \ref{ass:ODEf} and \ref{ass:ODEf2} and using \eqref{eq:storage} gives
\begin{align}\label{eq:Fout}
\begin{split}
F^{\mathrm{T}}(t)v(t) & \geq  \dot{V}\bigl(z(\cdot,t)\bigr)+ \omega\norm{z(\cdot,t)}_{L^2((0,1);\mathbb{R}^{2})}^2.
\end{split}
\end{align}
Integrating the above \eqref{eq:Fout} immediately yields \eqref{eq:passivity}. The extension to mild solutions may be worked out using standard convergence arguments. 
\end{proof}
\end{proposition}
The interpretation of Proposition \ref{prop:pass} above is the following: the output \eqref{eq:outputF} of the PDE subsystem represents the forces generated by the rolling contact process, which is essentially governed by friction, and therefore passive (for a discussion about passivity and dissipativity of distributed friction models, the reader may consult, e.g., \textcite{DistrLuGre,FrBD}). In the interconnection \eqref{eq:originalSystems}, these forces act, in turn, as the input to the ODE subsystem \eqref{eq:originalSystemsODE}. These peculiar features of the ODE-PDE coupling \eqref{eq:originalSystems} will be exploited in Section \ref{sect:Controllllll} to synthesize a stabilizing controller.
Clearly, the strict dissipativity inequality \eqref{eq:dissA0001} also ensures the existence of an inverse operator $\mathscr{A}_{\Sigma}^{-1}(y)$ on $L^2((0,1);\mathbb{R}^{2})$. Accordingly, for all $y \in \mathbb{R}^{2}$, the matrix $ \mathbf{M}_{2}(\mathbb{R}) \ni \Psi(y) \triangleq -(\mathscr{K}_1\mathscr{A}_{\Sigma}^{-1}(y)H)$ is introduced. Lengthy but straightforward manipulations show that $\Psi(y)$ is invertible for any combination of model parameters, that is, $\Psi(y) \in \mathbf{GL}_{2}(\mathbb{R})$ for all $y \in \mathbb{R}^2$. In this context, it is perhaps worth observing that, for all $y \in \mathbb{R}^{2}$, the existence of a unique matrix $\Psi(y)$, as defined bove, is ensured by the invertibility of the operator $(\mathscr{A}_{\Sigma}(y),\mathscr{D}(\mathscr{A}_{\Sigma}(y)))$. On the other hand, the invertibility of $\Psi(y)$ implies the following result, which is propaedeutic to synthesizing the proposed output-feedback stabilizing controller.

\begin{lemma}\label{prop:Prop1}
Suppose that Assumptions \ref{ass:ODEf} and \ref{ass:ODEf2} hold and consider the matrix $\mathbf{GL}_{2}(\mathbb{R}) \ni \Psi(y) \triangleq -(\mathscr{K}_1\mathscr{A}_{\Sigma}^{-1}(y)H)$. Then, for every $y \in \mathbb{R}^{2}$, there exists a unique solution $M(\cdot,y) \in C^1([0,1];\mathbf{M}_{2}(\mathbb{R}))$ to the nonlocal matrix ODE
\begin{subequations}\label{eq:ODEMatr}
\begin{align}
\begin{split}
\Lambda\dpd{M(\xi,y)}{\xi} &=\theta\Sigma(y)\bigl[M(\xi,y) + (\mathscr{K}_2M)(y)\bigr] +(\mathscr{K}_3M)(y) \\
& \quad + H\Psi^{-1}(y), \quad \xi \in (0,1), \label{eq:ODEMatr1}
\end{split}\\
M(0,y) &= 0,\label{eq:ODEMatr2}
\end{align}
\end{subequations}
satisfying the normalization condition $(\mathscr{K}_1M)(y) = I_{2}$.
\begin{proof}
Since $(\mathscr{A}_{\Sigma}(y),\mathscr{D}(\mathscr{A}_{\Sigma}(y)))$ generates an exponentially stable $C_0$-semigroup on $L^2((0,1);\mathbb{R}^{2})$, its spectrum cannot contain $0$, that is, $0 \notin \sigma(\mathscr{A}_{\Sigma}(y))$, implying that $(\mathscr{A}_{\Sigma}(y),\mathscr{D}(\mathscr{A}_{\Sigma}(y)))$ must be invertible on $L^2((0,1);\mathbb{R}^{2})$, and therefore also on $L^2((0,1);\mathbf{M}_{2}(\mathbb{R}))$.
Consequently, the unique solution $M(\xi,y)$ to the PDE \eqref{eq:ODEMatr} with BC \eqref{eq:ODEMatr2} is formally given by 
\begin{align}
M(\xi,y) = -(\mathscr{A}_{\Sigma}^{-1}(y)H)(\xi)\Psi^{-1}(y), \quad \xi \in [0,1].
\end{align}
It is easily verified that $M(\cdot,y) \in C^1([0,1];\mathbf{M}_{2}(\mathbb{R}))$. Moreover, computing $(\mathscr{K}_1M)(y)$ and recalling the definition of $\Psi(y)$ yields the identity $(\mathscr{K}_1M)(y) = I_{2}$.
\end{proof}
\end{lemma}
In the linear case ($\theta = 0$), the assertion of Lemma \ref{prop:Prop1} is equivalent to $\ran(G_1\Psi(\cdot)\cdot) = \mathbb{R}^{2}$. Conversely, in the semilinear one, Lemma \ref{prop:Prop1} merely states that the equilibria $(X^\star,z^\star) \in \mathbb{R}^{2}\times\mathscr{D}(\mathscr{A})$ associated with a constant input $U^\star \in \mathbb{R}^{2}$ solve the following nonlinear system: 
\begin{subequations}\label{eq:equilibria}
\begin{align}
\Psi^{-1}(v^\star)G_1^+(X^\star+b) +v^\star &= 0, \\
z^\star(\xi) +(\mathscr{A}_\Sigma^{-1}(v^\star)H)v^\star &= 0, \label{eq:uStatEqil} \\
v^\star -v(X^\star,U^\star) &= 0,
\end{align}
\end{subequations}
with $v^\star \in \mathbb{R}^{2}$.


\section{Controller design}\label{sect:Controllllll}
The present section is dedicated to synthesizing state and output-feedback backstepping controllers that exponentially stabilize \eqref{eq:originalSystems} around a desired equilibrium $(X^\star,z^\star) \in \mathbb{R}^{2}\times \mathscr{D}(\mathscr{A})$ as in \eqref{eq:equilibria}, corresponding to a stationary input $U(t) = U^\star$.

\subsection{State-feedback controller design}\label{sect:StateFedd}

To streamline the mathematical treatment, the variables $\mathbb{R}^{2} \ni X_\delta(t) \triangleq X(t)-X^\star$, $\mathbb{R}^{2} \ni z_\delta(\xi,t) \triangleq z(\xi,t) - z^\star(\xi)$, and $\mathbb{R}^{2} \ni U_\delta(t) \triangleq U(t)-U^\star$ are first introduced, so that the following system may be considered in place of the original one:
\begin{subequations}\label{eq:originalSystemsEqSF}
\begin{align}
\begin{split}
& \dot{X}_\delta(t) = A_1X_\delta(t) + G_1(\mathscr{K}_1z_\delta)(t), \quad t \in (0,T),
\end{split} \label{eq:originalSystemsODEEqSF}\\
\begin{split}
& \dpd{z_\delta(\xi,t)}{t} + \Lambda \dpd{z_\delta(\xi,t)}{\xi} =\theta\Sigma\bigl(v(X(t),U(t))\bigr) \\
& \qquad \times \bigl[z_\delta(\xi,t) + (\mathscr{K}_2z_\delta)(t)\bigr]   +(\mathscr{K}_3z_\delta)(t)\\
& \qquad + \theta\Bigl(\Sigma\bigl(v(X(t),U(t))\bigr)-\Sigma(v^\star)\Bigr) \bigl[z^\star(\xi) + (\mathscr{K}_2z^\star)\bigr] \\
& \qquad+Hv\bigl(X_\delta(t),U_\delta(t)\bigr),\quad   (\xi,t) \in (0,1) \times (0,T),
\end{split}  \label{eq:originalSystemsPDEEqSF} \\
& z_\delta(0,t) = 0, \quad t \in (0,T).\label{eq:originalSystemsBCEqSF}
\end{align}
\end{subequations}
Furthermore, to proceed with the design of an appropriate control law, the following auxiliary variable is defined:
\begin{align}\label{eq:Z1SF}
Z(t) & \triangleq (\mathscr{K}_1z_\delta)(t) - \varpi\bigl(X_\delta(t)\bigr),
\end{align}
where $\varpi(X_\delta(t)) \in \mathbb{R}^{2}$ represents a linear \emph{virtual control law}, whose particular expression needs yet to be specified. Finally, the following transformation is considered:
\begin{align}
\zeta(\xi,t) & \triangleq z_\delta(\xi,t)-M(\xi,v^\star)\varpi\bigl(X_\delta(t)\bigr), \label{eq:Z2_obSF}
\end{align}
where $M(\cdot,v^\star) \in C^1([0,1];\mathbf{M}_{2}(\mathbb{R}))$ is the matrix-valued function of Lemma \ref{prop:Prop1}. Accordingly, the entire strategy is then articulated into three main steps.

\subsubsection{Step 1}
Substituting \eqref{eq:Z1SF} into \eqref{eq:originalSystemsODEEqSF} provides
\begin{align}\label{eq:Z_1dotSF}
\begin{split}
\dot{X}_\delta(t) & = A_1X_\delta(t) + G_1\Bigl[Z(t) +\varpi\bigl(X_\delta(t)\bigr)\Bigr], \quad t \in (0,T).
\end{split}
\end{align}
Now, the following Lyapunov function candidate is considered:
\begin{align}\label{eq:Lyapunov1SF}
V_1\bigl(X_\delta(t)\bigr) \triangleq \dfrac{1}{2}X_\delta^{\mathrm{T}}(t)X_\delta(t).
\end{align}
Differentiating \eqref{eq:Lyapunov1SF} along the dynamics \eqref{eq:Z_1dotSF} yields
\begin{align}\label{eq:V1dot}
\begin{split}
\dot{V}_1(t) & = X_\delta^{\mathrm{T}}(t)\Bigl[A_1X_\delta(t)+G_1\varpi\bigl(X_\delta(t)\bigr)\Bigr] \\
& \quad + X_\delta^{\mathrm{T}}(t) G_1Z(t), \quad t\in(0,T).
\end{split}
\end{align}
Therefore, specifying the virtual control law as
\begin{align}\label{eq:alpha1SFSF}
\begin{split}
\varpi\bigl(X_\delta(t)\bigr)  & = -G_1^+A_1^* X_\delta(t),
\end{split}
\end{align}
where $\mathbf{M}_{2}(\mathbb{R}) \ni A_1^* \triangleq A_1 + qI_{2}$ for some appropriately chosen $q\in \mathbb{R}_{>0}$, gives
\begin{align}
\begin{split}
\dot{X}_\delta(t) & = -qX_\delta(t)+G_1(\mathscr{K}_1\zeta)(t), 
\end{split} \label{eq:Z1dotFinSF}\\
\begin{split}
\dot{V}_1(t) & \leq -q\norm{X_\delta(t)}_2^2 + X_\delta^{\mathrm{T}}(t)G_1Z(t), \quad t\in(0,T). \label{eq:SFVa} 
\end{split}
\end{align}
In preparation for the next step, the time derivative of the virtual control law is also computed explicitly:
\begin{align}\label{eq:dotAlphaSF}
\begin{split}
\dot{\varpi}\bigl(X_\delta(t),\zeta(\cdot,t)\bigr)  & = \dod{\varpi(X_\delta)}{X_{\delta}} \dot{X}_\delta(t)  = -G_1^+A_1^*\bigl[G_1Z(t)-qX_\delta(t)\bigr].
\end{split}
\end{align}
This concludes \textbf{Step 1}. 

\subsubsection{Step 2}\label{sect:step2}

By invoking Lemma \ref{prop:Prop1}, it is possible to transform the PDE subsystem dynamics into a target system whose exponential convergence also implies that of $Z(t)$.
\begin{lemma}
Under Assumptions \ref{ass:ODEf} and \ref{ass:ODEf2}, the transformation \eqref{eq:Z2_obSF}, with the matrix-valued function $M(\cdot,v^\star) \in C^1([0,1];\mathbf{M}_{2}(\mathbb{R}))$ as in Lemma \ref{prop:Prop1}, converts the PDE subsystem \eqref{eq:originalSystemsPDEEqSF}-\eqref{eq:originalSystemsBCEqSF} into
\begin{subequations}\label{eq:PDEzetaTot0SF}
\begin{align}\label{eq:PDEzeta0}
\begin{split}
& \dpd{\zeta(\xi,t)}{t} + \Lambda \dpd{\zeta(\xi,t)}{\xi} =\theta\Sigma\bigl(v(X(t),U(t))\bigr) \\
& \qquad \times \bigl[\zeta(\xi,t) + (\mathscr{K}_2\zeta)(t)\bigr] +(\mathscr{K}_3\zeta)(t)\\ 
& \qquad + \theta\Bigl(\Sigma\bigl(v(X(t),U(t))\bigr)-\Sigma(v^\star)\Bigr) \bigl[z^\star(\xi) + (\mathscr{K}_2z^\star)\bigr] \\
& \qquad +H\Bigl[v\bigl(X_\delta(t),U_\delta(t)\bigr)-\Psi^{-1}(v^\star)\varpi\bigl(X_\delta(t)\bigr)\Bigr] \\
& \qquad +\theta\Bigl(\Sigma\bigl(v(X(t),U(t))\bigr)-\Sigma(v^\star)\Bigr) \\
& \qquad \times \bigl[M(\xi,v^\star) + (\mathscr{K}_2M)(v^\star)\bigr]\varpi\bigl(X_\delta(t)\bigr) \\
& \qquad  -M(\xi,v^\star)\dot{\varpi}\bigl(X_\delta(t),\zeta(\cdot,t)\bigr), \quad (\xi,t) \in (0,1) \times (0,T),
\end{split}\\
& \zeta(0,t) = 0, \quad t \in (0,T),
\end{align}
\end{subequations}
with $(\mathscr{K}_1\zeta)(t) = Z(t)$.
\begin{proof}
The result follows from straightforward calculations and is therefore omitted for brevity.
\end{proof}
\end{lemma}
More conveniently, the new PDE subsystem \eqref{eq:PDEzetaTot0SF} permits to prove exponential stability in the spatial $L^2$-norm for the state $\zeta(\xi,t)$, which automatically implies that of $Z = (\mathscr{K}_1\zeta)(t)$. In order to derive the expression for the control law $U_\delta(t)$, the following Lyapunov function candidate is considered:
\begin{align}\label{eq:Lyapunov2SF}
V_2\bigl(\zeta(\cdot,t)\bigr)\triangleq \dfrac{1}{2} \int_0^1 \zeta^{\mathrm{T}}(\xi,t)\mathscr{Q}(\xi)\zeta(\xi,t) \dif \xi,
\end{align}
which coincides with the storage function of Proposition \ref{prop:pass}. 
Taking the derivative of \eqref{eq:Lyapunov2SF} along the dynamics \eqref{eq:PDEzetaTot0SF} yields
\begin{align}\label{eq:V_dder1SF}
\begin{split}
\dot{V}_2(t) & = \int_0^1\zeta^{\mathrm{T}}(\xi,t)\mathscr{Q}(\xi)(\mathscr{A}\zeta)(\xi) \dif \xi \\
& \quad +\theta\int_0^1 \zeta^{\mathrm{T}}(\xi,t)\mathscr{Q}(\xi)\Sigma\bigl(v(X(t),U(t))\bigr)\\
& \quad \times \bigl[\zeta(\xi,t) + (\mathscr{K}_2\zeta)(t)\bigr] \dif \xi \\
& \quad +  \theta\int_0^1 \zeta^{\mathrm{T}}(\xi,t)\mathscr{Q}(\xi)\Bigl(\Sigma\bigl(v(X(t),U(t))\bigr)-\Sigma(v^\star)\Bigr)\\
& \quad \times  \bigl[z^\star(\xi) + (\mathscr{K}_2z^\star)\bigr]\dif \xi  \\
& \quad +Z^{\mathrm{T}}(t)\Bigl[v\bigl(X_\delta(t),U_\delta(t)\bigr)-\Psi^{-1}(v^\star)\varpi\bigl(X_\delta(t)\bigr)\Bigr]  \\
& \quad +\theta \int_0^1 \zeta^{\mathrm{T}}(\xi,t)\mathscr{Q}(\xi)\Bigl(\Sigma\bigl(v(X(t),U(t))\bigr)-\Sigma(v^\star)\Bigr) \\
& \quad \times \bigl[M(\xi,v^\star) + (\mathscr{K}_2M)(v^\star)\bigr]\varpi\bigl(X_\delta(t)\bigr)\dif \xi \\
& \quad  -Z_M^{\mathrm{T}} (t,v^\star)\dot{\varpi}\bigl(X_\delta(t),\zeta(\cdot,t)\bigr), \quad t \in (0,T),
\end{split}
\end{align}
where, for convenience of notation,
\begin{align}
Z_M (t,v^\star) & = (\mathscr{K}_M \zeta)(t,v^\star) \triangleq \int_0^1 M^{\mathrm{T}}(\xi,v^\star)\mathscr{Q}(\xi)\zeta(\xi,t) \dif \xi,
\end{align}
has been introduced.

By Assumption \ref{ass:ODEf}, the sum of the first two terms appearing on the right-hand side of \eqref{eq:V_dder1SF} is negative definite. Moreover, recalling the definition of $\dot{\varpi}(X_\delta(t),\zeta(\cdot,t)) $ according to \eqref{eq:dotAlphaSF} provides
\begin{align}\label{eq:V2Intermediate}
\begin{split}
 \dot{V}_2(t) & \leq - \omega\norm{\zeta(\cdot,t)}_{L^2((0,1);\mathbb{R}^{2})}^2 \\
& \quad  +Z^{\mathrm{T}}(t)\Bigl[v\bigl(X_\delta(t),U_\delta(t)\bigr)-\Psi^{-1}(v^\star)\varpi\bigl(X_\delta(t)\bigr)\Bigr]  \\
& \quad +Z_M^{\mathrm{T}} (t,v^\star) G_1^+A_1^*\bigl[G_1Z(t) -qX_\delta(t)\bigr],  \\
& \quad +  \theta\int_0^1 \zeta^{\mathrm{T}}(\xi,t)\mathscr{Q}(\xi)\Bigl(\Sigma\bigl(v(X(t),U(t))\bigr)-\Sigma(v^\star)\Bigr)\\
& \quad \times  \bigl[z^\star(\xi) + (\mathscr{K}_2z^\star)\bigr]\dif \xi \\
& \quad + \theta\int_0^1 \zeta^{\mathrm{T}}(\xi,t)\mathscr{Q}(\xi)\Bigl(\Sigma\bigl(v(X(t),U(t))\bigr)-\Sigma(v^\star)\Bigr) \\
& \quad \times \bigl[M(\xi,v^\star) + (\mathscr{K}_2M)(v^\star)\bigr]\varpi\bigl(X_\delta(t)\bigr)\dif \xi, \quad t \in (0,T). 
\end{split}
\end{align}
Hence, the second part of the control input is selected as
\begin{align}\label{eq:U_deltaSF}
\begin{split}
U_\delta(t) & = -G_2^+\Bigl[\bigl(G_1^+A_1^*G_1\bigr)^{\mathrm{T}}Z_M(t,v^\star) +\gamma_1 G_1^{\mathrm{T}}X_\delta(t)\Bigr] \\
& \quad -G_2^+\Bigl[A_2X_\delta(t)- \Psi^{-1}(v^\star)\varpi\bigl(X_\delta(t)\bigr)\Bigr],
\end{split}
\end{align}
where the term $-\gamma_1G_2^+G_1^{\mathrm{T}}X_\delta(t)$, with $\gamma_1\in \mathbb{R}_{>0}$ to be specified later, is designed to eliminate the residual coupling $X_\delta^{\mathrm{T}}(t)G_1Z(t)$ appearing in \eqref{eq:SFVa}, and $ -G_2^+(G_1^+A_1^*G_1)^{\mathrm{T}}Z_M(t,v^\star)$ to suppress $Z_M^{\mathrm{T}}(t,v^\star)G_1^+A_1^*G_1Z(t)$ in \eqref{eq:V2Intermediate}. Inserting \eqref{eq:U_deltaSF} into \eqref{eq:PDEzetaTot0SF} and \eqref{eq:V2Intermediate} gives
\begin{subequations}\label{eq:PDEzetaTot2SF}
\begin{align}\label{eq:PDEzeta2SF}
\begin{split}
& \dpd{\zeta(\xi,t)}{t} + \Lambda \dpd{\zeta(\xi,t)}{\xi} =\theta\Sigma\bigl(v(X_\delta(t)+X^\star,U_\delta(t)+U^\star)\bigr) \\
& \qquad \times \bigl[\zeta(\xi,t) + (\mathscr{K}_2\zeta)(t)\bigr] +(\mathscr{K}_3\zeta)(t)\\ 
& \qquad -H\Bigl[\bigl(G_1^+A_1^*G_1\bigr)^{\mathrm{T}}(\mathscr{K}_M \zeta)(t,v^\star) +\gamma_1 G_1^{\mathrm{T}}X_\delta(t)\Bigr] \\
& \qquad + \theta\Bigl(\Sigma\bigl(v(X_\delta(t)+X^\star,U_\delta(t)+U^\star)\bigr)-\Sigma(v^\star)\Bigr) \\
& \qquad \times  \bigl[z^\star(\xi) + (\mathscr{K}_2z^\star)\bigr] \\
& \qquad +\theta \Bigl(\Sigma\bigl(v(X_\delta(t)+X^\star,U_\delta(t)+U^\star)\bigr)-\Sigma(v^\star)\Bigr) \\
& \qquad \times \bigl[M(\xi,v^\star) + (\mathscr{K}_2M)(v^\star)\bigr]\varpi\bigl(X_\delta(t)\bigr) \\
& \qquad  -M(\xi,v^\star)\dot{\varpi}\bigl(X_\delta(t),\zeta(\cdot,t)\bigr), \quad (\xi,t) \in (0,1) \times (0,T),
\end{split}\\
& \zeta(0,t) = 0, \quad t \in (0,T),
\end{align} 
\end{subequations}
and
\begin{align}
\begin{split}
  \dot{V}_2(t)  & \leq - \omega\norm{\zeta(\cdot,t)}_{L^2((0,1);\mathbb{R}^{2})}^2  -qZ_M^{\mathrm{T}} (t,v^\star) G_1^+A_1^*X_\delta(t)\\
& \quad +\cancel{Z_M^{\mathrm{T}} (t,v^\star) G_1^+A_1^*G_1Z(t)}-\cancel{Z^{\mathrm{T}}(t)\bigl(G_1^+A_1^*G_1\bigr)^{\mathrm{T}}Z_M(t,v^\star)} \\
& \quad -\gamma_1 Z^{\mathrm{T}}(t)G_1^{\mathrm{T}}X_\delta(t) \\
& \quad +  \theta\int_0^1 \zeta^{\mathrm{T}}(\xi,t)\mathscr{Q}(\xi)\Bigl(\Sigma\bigl(v(X(t),U(t))\bigr)-\Sigma(v^\star)\Bigr)\\
& \quad \times  \bigl[z^\star(\xi) + (\mathscr{K}_2z^\star)\bigr]\dif \xi \\
& \quad + \theta\int_0^1 \zeta^{\mathrm{T}}(\xi,t)\mathscr{Q}(\xi)\Bigl(\Sigma\bigl(v(X(t),U(t))\bigr)-\Sigma(v^\star)\Bigr) \\
& \quad \times \bigl[M(\xi,v^\star) + (\mathscr{K}_2M)(v^\star)\bigr]\varpi\bigl(X_\delta(t)\bigr)\dif \xi, \quad t \in (0,T).  \label{eq:V_2dot_obsCntrSF}
\end{split}
\end{align}
This concludes \textbf{Step 2}.

\subsubsection{Step 3}\label{sect:Step3}
The complete Lyapunov function is finally assembled as
\begin{align}\label{eq:LyapunovFinalSF}
\begin{split}
& V\bigl(X_\delta(t),\zeta(\cdot,t)\bigr)  \triangleq V_1\bigl(X_\delta(t)\bigr)+ \dfrac{1}{\gamma_1}V_2\bigl(\zeta(\cdot,t)\bigr).
\end{split}
\end{align}
Taking the derivative of \eqref{eq:LyapunovFinalSF} yields
\begin{align}\label{eq:V_3iiSF}
\begin{split}
\dot{V}(t) & \leq -q\norm{X(t)}_2^2+\cancel{X_\delta^{\mathrm{T}}(t)G_1Z(t)}-\dfrac{\omega}{\gamma_1}\norm{\zeta(\cdot,t)}_{L^2((0,1);\mathbb{R}^{2})}^2\\
& \quad- \cancel{Z^{\mathrm{T}}(t)G_1^{\mathrm{T}}X_\delta(t)}-\dfrac{q}{\gamma_1}Z_M^{\mathrm{T}}(t,v^\star)G_1^+A_1^*X_\delta(t)\\
& \quad +  \dfrac{\theta}{\gamma_1}\int_0^1 \zeta^{\mathrm{T}}(\xi,t)\mathscr{Q}(\xi)\Bigl(\Sigma\bigl(v(X(t),U(t))\bigr)-\Sigma(v^\star)\Bigr)\\
& \quad \times  \bigl[z^\star(\xi) + (\mathscr{K}_2z^\star)\bigr]\dif \xi \\
& \quad + \dfrac{\theta}{\gamma_1}\int_0^1 \zeta^{\mathrm{T}}(\xi,t)\mathscr{Q}(\xi)\Bigl(\Sigma\bigl(v(X(t),U(t))\bigr)-\Sigma(v^\star)\Bigr) \\
& \quad \times \bigl[M(\xi,v^\star) + (\mathscr{K}_2M)(v^\star)\bigr]\varpi\bigl(X_\delta(t)\bigr)\dif \xi, \quad t \in (0,T). 
\end{split}
\end{align}
The first cross term remaining in \eqref{eq:V_3iiSF} may be bounded by observing that
\begin{align}
\norm{Z_M(t,v^\star)}_2 & \leq \norm{M^{\mathrm{T}}(\cdot,v^\star)\mathscr{Q}(\cdot)}_\infty\norm{\zeta(\cdot,t)}_{L^2((0,1);\mathbb{R}^{2})}.
\end{align}
Thus, applying Cauchy-Schwarz and then the generalized Young's inequality for products provides
\begin{align}
\begin{split}
& qZ_M^{\mathrm{T}}(t,v^\star)G_1^+A_1^*X(t)  \leq \dfrac{q^2}{2\varepsilon}\norm{G_1^+A_1^*}^2 \norm{X(t)}_2^2 \\
& \qquad + \dfrac{\varepsilon}{2}\norm{M^{\mathrm{T}}(\cdot,v^\star)\mathscr{Q}(\cdot)}_\infty^2\norm{\zeta(\cdot,t)}_{L^2((0,1);\mathbb{R}^{2})}^2,
\end{split}
\end{align}
for $\varepsilon \in \mathbb{R}_{>0}$ to be appropriately chosen. Specifically, by setting
\begin{align}
\varepsilon &  =\varepsilon(v^\star) \triangleq \dfrac{\omega}{\norm{M^{\mathrm{T}}(\cdot,v^\star)\mathscr{Q}(\cdot)}_\infty^2},  \nonumber \\
\begin{split}
\gamma_1 & = \dfrac{q}{\omega}\norm{G_1^+A_1^*}^2\norm{M^{\mathrm{T}}(\cdot,v^\star)\mathscr{Q}(\cdot)}_\infty^2, \label{eq:gamma_1}
\end{split}
\end{align}
gives
\begin{align}\label{eq:VsemidiNNND}
\begin{split}
& \dot{V}(t)  \leq -\gamma V(t) \\
& \quad +  \dfrac{\theta}{\gamma_1(v^\star)}\int_0^1 \zeta^{\mathrm{T}}(\xi,t)\mathscr{Q}(\xi)\Bigl(\Sigma\bigl(v(X(t),U(t))\bigr)-\Sigma(v^\star)\Bigr)\\
& \quad \times  \bigl[z^\star(\xi) + (\mathscr{K}_2z^\star)\bigr]\dif \xi \\
& \quad + \dfrac{\theta}{\gamma_1(v^\star)}\int_0^1 \zeta^{\mathrm{T}}(\xi,t)\mathscr{Q}(\xi)\Bigl(\Sigma\bigl(v(X(t),U(t))\bigr)-\Sigma(v^\star)\Bigr) \\
& \quad \times \bigl[M(\xi,v^\star) + (\mathscr{K}_2M)(v^\star)\bigr]\varpi\bigl(X_\delta(t)\bigr)\dif \xi, \quad t \in (0,T),
\end{split}
\end{align}
with $\mathbb{R}_{>0}\ni \gamma \triangleq \min\{q, \frac{\omega}{ \norm{\lambda\ped{max}(\mathscr{Q}(\cdot))}_\infty}\}$, where $\mathbb{R}_{>0} \ni \lambda\ped{max}(\mathscr{Q}(\xi))$ denotes the largest eigenvalue of $\mathscr{Q}(\xi)$.
Moreover, by noting that 
\begin{subequations}
\begin{align}
\begin{split}
& \int_0^1 \zeta^{\mathrm{T}}(\xi,t)\mathscr{Q}(\xi)\Bigl(\Sigma\bigl(v(X(t),U(t))\bigr)-\Sigma(v^\star)\Bigr)\\
& \quad \times \bigl[z^\star(\xi) + (\mathscr{K}_2z^\star)\bigr]\dif \xi\\
& \quad  \leq L_\Sigma\norm{\mathscr{Q}(\cdot)}_\infty\norm{z^\star(\cdot) + (\mathscr{K}_2z^\star)}_{L^2((0,1);\mathbb{R}^{2})} \\
& \qquad \times \norm{\zeta(\cdot,t)}_{L^2((0,1);\mathbb{R}^{2})}\norm{A_2X_\delta(t)+G_2U_\delta(t)}_2, 
\end{split} \\
\begin{split}
& \int_0^1 \zeta^{\mathrm{T}}(\xi,t)\mathscr{Q}(\xi)\Bigl(\Sigma\bigl(v(X(t),U(t))\bigr)-\Sigma(v^\star)\Bigr) \\
& \quad \times \bigl[M(\xi,v^\star) + (\mathscr{K}_2M)(v^\star)\bigr]\varpi\bigl(X_\delta(t)\bigr)\dif \xi \\
& \quad \leq L_\Sigma\norm{\mathscr{Q}(\cdot)}_\infty\norm{M(\cdot,v^\star) + (\mathscr{K}_2M)(v^\star)}_\infty  \\
& \qquad \times \norm{\zeta(\cdot,t)}_{L^2((0,1);\mathbb{R}^{2})}\\
& \qquad \times\norm{A_2X_\delta(t)+G_2U_\delta(t)}_2\norm{G_1^+A_1^*X_\delta(t)}_2,
\end{split}
\end{align}
\end{subequations}
it is possible to infer the existence of $\eta_1(v^\star),\eta_2(v^\star) \in \mathbb{R}_{\geq 0}$ such that
\begin{subequations}\label{eq:inferrEta1Eta2}
\begin{align}
\begin{split}
& \dfrac{1}{\gamma_1(v^\star)}\int_0^1 \zeta^{\mathrm{T}}(\xi,t)\mathscr{Q}(\xi)\Bigl(\Sigma\bigl(v(X(t),U(t))\bigr)-\Sigma(v^\star)\Bigr)\\
& \quad \times\bigl[z^\star(\xi) + (\mathscr{K}_2z^\star)\bigr]\dif \xi \leq \eta_1(v^\star)V(t), 
\end{split} \\
\begin{split}
&\dfrac{1}{\gamma_1(v^\star)} \int_0^1 \zeta^{\mathrm{T}}(\xi,t)\mathscr{Q}(\xi)\Bigl(\Sigma\bigl(v(X(t),U(t))\bigr)-\Sigma(v^\star)\Bigr) \\
& \quad \times \bigl[M(\xi,v^\star) + (\mathscr{K}_2M)(v^\star)\bigr]\varpi\bigl(X_\delta(t)\bigr)\dif \xi  \leq \eta_2(v^\star)V^{\frac{3}{2}}(t),
\end{split}
\end{align}
\end{subequations}
where, from the definition of the equilibrium \eqref{eq:uStatEqil} with $z^\star(\xi) = -(\mathscr{A}_\Sigma^{-1}(v^\star)H)v^\star$, it follows that $\eta_1(0) = 0$. In turn, inserting \eqref{eq:inferrEta1Eta2} into \eqref{eq:VsemidiNNND} provides
\begin{align}\label{eq:VfinDerDynSF}
\dot{V}(t) \leq - \bigl(\gamma-\theta\eta_1(v^\star)\bigr) V(t) + \theta\eta_2(v^\star)V^{\frac{3}{2}}(t), \quad t \in(0,T).
\end{align}
Theorem \ref{theorem:final} asserts the main result of the paper.
\begin{theorem}\label{theorem:final}
Consider the ODE-PDE interconnection \eqref{eq:originalSystems}-\eqref{eq:MatricesOOO} under Assumptions \ref{ass:ODEf} and \ref{ass:ODEf2}, along with the control law $U(t) = U^\star + U_\delta(t)$, with $U_\delta(t)$ as in \eqref{eq:U_deltaSF}, and suppose that the target equilibrium $(X^\star,z^\star) \in \mathbb{R}^{2}\times \mathscr{D}(\mathscr{A})$ corresponding to the input $U^\star \in \mathbb{R}^{2}$ is such that $\mathbb{R}^{2} \ni v^\star = A_2X^\star + G_2U^\star$ verifies $\theta\eta_1(v^\star) < \gamma$. Then, for all ICs $(X_0,z_0) \in \mathcal{X}$ such that
\begin{align}\label{eq:V0_ini}
\begin{split}
V(0) & < \biggl(\dfrac{\gamma-\theta\eta_1(v^\star)}{\theta\eta_2(v^\star)}\biggr)^2,
\end{split}
\end{align}
with $V(X_\delta(t),\zeta(\cdot,t))$ as in \eqref{eq:LyapunovFinalSF}, the system \eqref{eq:Z1dotFinSF} and \eqref{eq:PDEzetaTot2SF} admits a unique mild solution $(X,z) \in C^0(\mathbb{R}_{\geq 0};\mathcal{X})$ satisfying
\begin{align}\label{eq:boundPErf}
\begin{split}
& \norm{(X(t)-X^\star,z(\cdot,t)-z^\star(\cdot))}_{\mathcal{X}} \\
& \quad \leq \beta(v^\star)\eu^{-\sigma t} \norm{(X_0-X^\star,z_0(\cdot)-z^\star(\cdot))}_{\mathcal{X}}, \quad t\in[0,T],
\end{split}
\end{align}
for some $\beta(v^\star), \sigma \in \mathbb{R}_{>0}$.
\begin{proof}
Recalling the global Lipschitz condition \eqref{eq:SigmaLSigma} introduced in Assumption \ref{ass:ODEf2}, and observing that the control input $U_\delta(t)$ in \eqref{eq:U_deltaSF} only contains bounded functionals of $\zeta(\xi,t)$, it follows from standard semigroup arguments for semilinear problems (see, e.g., Theorem 11.1.5 in \textcite{Zwart} or 6.1.4 in \textcite{Pazy}) that the closed loop ODE-PDE interconnection described by \eqref{eq:Z1dotFinSF} and \eqref{eq:PDEzetaTot2SF} admits a unique local mild solution $(X_\delta,\zeta) \in C^0([0,t\ped{max});\mathcal{X})$ for all ICs $(X_{\delta,0},\zeta_0)\triangleq (X_\delta(0),\zeta(\cdot,0)) \in \mathcal{X}$. Consequently, from the transformation \eqref{eq:Z2_obSF} and the fact that $X(t) \triangleq X_\delta(t) +X^\star$ and $z(\xi,t) \triangleq z_\delta(\xi,t) + z^\star(\xi)$, with $z^\star \in \mathscr{D}(\mathscr{A})$, it may be concluded the original ODE-PDE interconnection \eqref{eq:originalSystems} also admits a unique mild solution $(X,z) \in C^0([0,t\ped{max});\mathcal{X})$ for all ICs $(X_0,z_0)\in \mathcal{X}$.
Moreover, according to Theorem \ref{thm:mild}, to prove global well-posedness, it is sufficient to show that $\norm{(X(t),z(\cdot,t))}_{\mathcal{X}} < \infty$ for all $t \in \mathbb{R}_{\geq 0}$, which is implied by inequality \eqref{eq:boundPErf}. In particular, for sufficiently regular solutions, the condition $\theta\eta_1(v^\star) < \gamma$, in conjunction with the bound \eqref{eq:V0_ini}, ensures the existence of a constant $\mu\in \mathbb{R}_{>0}$ such that
\begin{align}\label{eq:V0_ini22}
\dot{V}(t) \leq -\mu V(t), \quad t \in (0,T).
\end{align}
Thus, from an application of Grönwall-Bellman's inequality and the fact that the Lyapunov function $V(X_\delta(t),\zeta(\cdot,t))$ is equivalent to the squared norm $\norm{(X_\delta(t),\zeta(\cdot,t))}_{\mathcal{X}}^2$ on $\mathcal{X}$, the existence of $\beta_1, \sigma \in \mathbb{R}_{>0}$ may be inferred such that
\begin{align}\label{eq:XNormksks}
\norm{(X_\delta(t),\zeta(\cdot,t))}_{\mathcal{X}} \leq \beta_1\eu^{-\sigma t}\norm{(X_{\delta,0},\zeta_{0}(\cdot))}_{\mathcal{X}}, \quad t \in [0,T].
\end{align}
Recalling \eqref{eq:Z2_obSF} and \eqref{eq:alpha1SFSF}, and using the triangle inequality also yields
\begin{align}\label{eq:boundUdelta}
\begin{split}
& \norm{z_\delta(\cdot,t)}_{L^2((0,1);\mathbb{R}^n)} \\
& \quad \leq \norm{\zeta(\cdot,t)}_{L^2((0,1);\mathbb{R}^n)} + \norm{M(\cdot,v^\star)G_1^+A_1^*}_\infty\norm{X_\delta(t)}_2 \\
& \quad \leq \beta_2(v^\star)\eu^{-\sigma t}\norm{(X_{\delta,0},\zeta_{0}(\cdot))}_{\mathcal{X}}, \quad t\in[0,T],
\end{split}
\end{align}
with $\mathbb{R}_{>0}\ni \beta_2(v^\star) \triangleq \beta_1(1+ \norm{M(\cdot,v^\star)G_1^+A_1^*}_\infty)$.
Thus, combining \eqref{eq:XNormksks} and \eqref{eq:boundUdelta} provides
\begin{align}
\norm{(X_\delta(t),z_\delta(t))}_{\mathcal{X}} \leq \beta_2(v^\star)\eu^{-\sigma t}\norm{(X_{\delta,0},\zeta_{0}(\cdot))}_{\mathcal{X}}, \quad t\in[0,T],
\end{align}
which, in turn, implies \eqref{eq:boundPErf}. The result may then be extended to mild solutions using standard density arguments.
\end{proof}
\end{theorem}
Before moving to the synthesis of an output-feedback controller, some considerations are in order.
\begin{remark}
Assumptions \ref{ass:ODEf} and \ref{ass:ODEf2} are required to eliminate the cross terms $X_\delta^{\mathrm{T}}(t)G_1Z(t)$ and $Z_M^{\mathrm{T}}(t,v^\star)G_1^+A_1^*G_1Z(t)$ appearing in \eqref{eq:SFVa} and \eqref{eq:V2Intermediate}, respectively. Different choices of positive operators than $\mathscr{Q}(\xi)$ might preclude their direct suppression, introducing additional coupling sources between $X_\delta(t)$ and $\zeta(\xi,t)$. 
\end{remark}
\begin{remark}
The conditions prescribed by Theorem \ref{theorem:final} on the parameters $\eta_1(v^\star)$ and $\eta_2(v^\star)$ are clearly conservative. In essence, Theorem \ref{theorem:final} states that, for sufficiently small target equilibria and initial conditions, convergence towards the desired equilibrium will be achieved exponentially fast. In particular, it is easy to verify that the inequality $\theta\eta_1(v^\star) < \gamma$ is identically fulfilled for the zero input $U^\star =0$ whenever $b = 0$. Additionally, in the linear case ($\theta = 0$), Theorem \ref{theorem:final} asserts global well-posedness and uniform exponential stability around any equilibrium $(X^\star,z^\star) \in \mathbb{R}^2 \times \mathscr{D}(\mathscr{A})$: the estimate \eqref{eq:boundPErf} holds for all ICs $(X_0,z_0) \in \mathcal{X}$.
\end{remark}

\subsection{Output-feedback controller design}\label{sect:OutFedd}
Next, an output-feedback stabilizing controller is synthesized. To this end, it may be first noted that $(A_1,HA_2)$ is an observable pair for all combinations of model parameters.
Thus, denoting the estimates of $X(t)$, $z(\xi,t)$, and $Y(t)$ respectively as $\hat{X}(t) \in \mathbb{R}^{2}$, $\hat{z}(\xi,t) \in \mathbb{R}^{2}$, and $\hat{Y}(t) \in \mathbb{R}^{2}$, the following observer structure is proposed:
\begin{subequations}\label{eq:originalSystemsObs}
\begin{align}
\begin{split}
& \dot{\hat{X}}(t) =A_1\hat{X}(t)+ G_1(\mathscr{K}_1\hat{z})(t)\\
& \qquad \quad +b-L_1\bigl(Y(t)-\hat{Y}(t)\bigr), \quad t \in (0,T),
\end{split} \label{eq:originalSystemsODEObs}\\
\begin{split}
& \dpd{\hat{z}(\xi,t)}{t} + \Lambda \dpd{\hat{z}(\xi,t)}{\xi} =\theta \Sigma\bigl(H^{-1}Y(t)\bigr)\bigl[\hat{z}(\xi,t) + (\mathscr{K}_2\hat{z})(t)\bigr]  \\
& \qquad \qquad  \qquad \qquad +(\mathscr{K}_3\hat{z})(t) +Y(t),\\
& \qquad \qquad\qquad\qquad  \quad (\xi,t) \in (0,1) \times (0,T),
\end{split} \label{eq:originalSystemsPDEObs} \\
& \hat{z}(0,t) = 0, \quad t \in (0,T),\label{eq:originalSystemsBCObs}
\end{align}
\end{subequations}
where $L_1 \in \mathbf{M}_{2\times 2}(\mathbb{R})$ is a matrix with constant coefficients. The estimated output reads
\begin{align}
\hat{Y}(t) = Hv\bigl(\hat{X}(t),U(t)\bigr) = H\bigl(A_2\hat{X}(t) + G_2U(t)\bigr).
\end{align}
Defining the error variables as $\mathbb{R}^{2} \ni \tilde{X}(t) \triangleq X(t)-\hat{X}(t)$ and $\mathbb{R}^{2}\ni \tilde{z}(\xi,t) \triangleq z(\xi,t)-\hat{z}(\xi,t)$.
\begin{lemma}[Existence of a Lyapunov function for the observer error dynamics]
Suppose that Assumptions \ref{ass:ODEf} and \ref{ass:ODEf2} hold. Then, there exist a matrix $\mathbf{Sym}_{2}(\mathbb{R}) \ni P \succ 0$ and constants $\phi, \rho \in \mathbb{R}_{>0}$ independent of $X(t)$ and $U(t)$ such that the Lyapunov function
\begin{align}
\begin{split}
V_0\bigl(\tilde{X}(t),\tilde{z}(\cdot,t)\bigr) & \triangleq \dfrac{1}{2}\tilde{X}^{\mathrm{T}}(t)P\tilde{X}(t) \\
& \quad + \dfrac{\phi}{2}\int_0^1 \tilde{z}^{\mathrm{T}}(\xi,t)\mathscr{Q}(\xi)\tilde{z}(\xi,t)\dif \xi
\end{split}
\end{align}
satisfies
\begin{align}
\dot{V}_0(t) \leq -\rho V_0(t), \quad t \in (0,T),
\end{align}
for all $\tilde{z}_0 \triangleq \tilde{z}(\cdot,0) \in \mathscr{D}(\mathscr{A})$.
\begin{proof}
The result follows from the fact that $L_1$ may be chosen such that $\mathbf{M}_{2}(\mathbb{R}) \ni \bar{A}_1 \triangleq A_1 + L_1HA_2$ is Hurwitz.
\end{proof}
\end{lemma}

In the output-feedback case, the second term of the control input, corresponding to \eqref{eq:U_deltaSF}, becomes
\begin{align}\label{eq:U_deltaOF}
\begin{split}
U_\delta(t) & = -G_2^+\Bigl[\bigl(G_1^+A_1^*G_1\bigr)^{\mathrm{T}}\hat{Z}_M(t,v^\star) +\gamma_1 G_1^{\mathrm{T}}\hat{X}_\delta(t)\Bigr] \\
& \quad -G_2^+\Bigl[A_2\hat{X}_\delta(t)- \Psi^{-1}(v^\star)\varpi\bigl(\hat{X}_\delta(t)\bigr)\Bigr],
\end{split}
\end{align} 
where $\hat{X}_\delta(t) \in \mathbb{R}^{2}$ and $ \mathbb{R}^{2} \ni \hat{Z}_M(t,v^\star) = (\mathscr{K}_M\hat{\zeta})(t)$ denote the estimates of $X_\delta(t)$ and $Z_M(t)$ constructed using the observer \eqref{eq:originalSystemsObs}. Utilizing \eqref{eq:U_deltaOF}, \textbf{Step 1} remains formally unchanged, yielding again \eqref{eq:Z1dotFinSF} and \eqref{eq:SFVa} for the ODE dynamics. On the other hand, the following equations may be derived governing the PDE and associated Lyapunov function dynamics:
\begin{subequations}\label{eq:PDEzetaTot2OF}
\begin{align}\label{eq:PDEzeta2OF}
\begin{split}
& \dpd{\zeta(\xi,t)}{t} + \Lambda \dpd{\zeta(\xi,t)}{\xi} =\theta\Sigma\bigl(v(X_\delta(t)+X^\star,U_\delta(t)+U^\star)\bigr) \\
& \qquad \times \bigl[\zeta(\xi,t) + (\mathscr{K}_2\zeta)(t)\bigr] +(\mathscr{K}_3\zeta)(t)\\ 
& \qquad +H\Bigl[A_2\tilde{X}(t)-\Psi^{-1}(v^\star)\varpi\bigl(\tilde{X}(t)\bigr)\Bigr]\\
& \qquad -H\Bigl[\bigl(G_1^+A_1^*G_1\bigr)^{\mathrm{T}}(\mathscr{K}_M \zeta)(t,v^\star) +\gamma_1 G_1^{\mathrm{T}}X_\delta(t)\Bigr] \\
& \qquad +H\Bigl[\bigl(G_1^+A_1^*G_1\bigr)^{\mathrm{T}}(\mathscr{K}_M \tilde{z})(t,v^\star) +\gamma_1 G_1^{\mathrm{T}}\tilde{X}(t)\Bigr] \\
& \qquad -H\bigl(G_1^+A_1^*G_1\bigr)^{\mathrm{T}}(\mathscr{K}_M M)(v^\star)\varpi\bigl(\tilde{X}(t)\bigr) \\
& \qquad +\theta \Bigl(\Sigma\bigl(v(X_\delta(t)+X^\star,U_\delta(t)+U^\star)\bigr)-\Sigma(v^\star)\Bigr) \\
& \qquad \times  \bigl[z^\star(\xi) + (\mathscr{K}_2z^\star)\bigr] \\
& \qquad +\theta \Bigl(\Sigma\bigl(v(X_\delta(t)+X^\star,U_\delta(t)+U^\star)\bigr)-\Sigma(v^\star)\Bigr) \\
& \qquad \times \bigl[M(\xi,v^\star) + (\mathscr{K}_2M)(v^\star)\bigr]\varpi\bigl(X_\delta(t)\bigr) \\
& \qquad  -M(\xi,v^\star)\dot{\varpi}\bigl(X_\delta(t),\zeta(\cdot,t)\bigr), \quad (\xi,t) \in (0,1) \times (0,T),
\end{split}\\
& \zeta(0,t) = 0, \quad t \in (0,T),
\end{align} 
\end{subequations}
and
\begin{align}
\begin{split}
  \dot{V}_2(t)  & \leq - \omega\norm{\zeta(\cdot,t)}_{L^2((0,1);\mathbb{R}^{2})}^2  -qZ_M^{\mathrm{T}} (t,v^\star) G_1^+A_1^*X_\delta(t)\\
& \quad -\gamma_1 Z^{\mathrm{T}}(t)G_1^{\mathrm{T}}X_\delta(t) \\
& \quad + \theta \int_0^1 \zeta^{\mathrm{T}}(\xi,t)\mathscr{Q}(\xi)\Bigl(\Sigma\bigl(v(X(t),U(t))\bigr)-\Sigma(v^\star)\Bigr)\\
& \quad \times  \bigl[z^\star(\xi) + (\mathscr{K}_2z^\star)\bigr]\dif \xi \\
& \quad +\theta \int_0^1 \zeta^{\mathrm{T}}(\xi,t)\mathscr{Q}(\xi)\Bigl(\Sigma\bigl(v(X(t),U(t))\bigr)-\Sigma(v^\star)\Bigr) \\
& \quad \times \bigl[M(\xi,v^\star) + (\mathscr{K}_2M)(v^\star)\bigr]\varpi\bigl(X_\delta(t)\bigr)\dif \xi \\
& \quad + Z^{\mathrm{T}}(t)\Bigl[A_2\tilde{X}(t)-\Psi^{-1}(v^\star)\varpi\bigl(\tilde{X}(t)\bigr)\Bigr] \\
& \quad +Z^{\mathrm{T}}(t)\Bigl[\bigl(G_1^+A_1^*G_1\bigr)^{\mathrm{T}}(\mathscr{K}_M \tilde{z})(t,v^\star) +\gamma_1 G_1^{\mathrm{T}}\tilde{X}(t)\Bigr] \\
& \quad -Z^{\mathrm{T}}\bigl(G_1^+A_1^*G_1\bigr)^{\mathrm{T}}(\mathscr{K}_M M)(v^\star)\varpi\bigl(\tilde{X}(t)\bigr) ,  \quad t \in (0,T).  \label{eq:V_2dot_obsCntrOF}
\end{split}
\end{align}
The complete Lyapunov function is finally assembled as
\begin{align}\label{eq:LyapunovFinal}
\begin{split}
& V\bigl(X_\delta(t),\zeta(\cdot,t),\tilde{X}(t),\tilde{z}(\cdot,t)\bigr) \triangleq \gamma_0V_0\bigl(\tilde{X}(t),\tilde{z}(\cdot,t)\bigr) \\
& \qquad \qquad + V_1\bigl(X_\delta(t)\bigr) + \dfrac{1}{\gamma_1}V_2\bigl(\zeta(\cdot,t)\bigr), 
\end{split}
\end{align}
where $ V_1(X_\delta(t))$ and $V_2(\zeta(\cdot,t))$ read as in \eqref{eq:Lyapunov1SF} and \eqref{eq:Lyapunov2SF}, respectively, and $\gamma_0 \in \mathbb{R}_{>0}$ is a constant to be appropriately selected. Taking the derivative of \eqref{eq:V_2dot_obsCntrOF} yields
\begin{align}\label{eq:finalDer}
\begin{split}
 & \dot{V}(t)   \leq -\gamma_0\rho V_0(t) - q\norm{X_\delta(t)}_2^2 - \dfrac{\omega}{\gamma_1}\norm{\zeta(\cdot,t)}_{L^2((0,1);\mathbb{R}^{2})}^2\\
& \quad  -\dfrac{q}{\gamma_1}Z_M^{\mathrm{T}} (t,v^\star) G_1^+A_1^*X_\delta(t)\\
& \quad + \dfrac{\theta}{\gamma_1} \int_0^1 \zeta^{\mathrm{T}}(\xi,t)\mathscr{Q}(\xi)\Bigl(\Sigma\bigl(v(X(t),U(t))\bigr)-\Sigma(v^\star)\Bigr)\\
& \quad \times  \bigl[z^\star(\xi) + (\mathscr{K}_2z^\star)\bigr]\dif \xi \\
& \quad +\dfrac{\theta}{\gamma_1} \int_0^1 \zeta^{\mathrm{T}}(\xi,t)\mathscr{Q}(\xi)\Bigl(\Sigma\bigl(v(X(t),U(t))\bigr)-\Sigma(v^\star)\Bigr) \\
& \quad \times \bigl[M(\xi,v^\star) + (\mathscr{K}_2M)(v^\star)\bigr]\varpi\bigl(X_\delta(t)\bigr)\dif \xi \\
& \quad + \dfrac{1}{\gamma_1}Z^{\mathrm{T}}(t)\Bigl[A_2\tilde{X}(t)-\Psi^{-1}(v^\star)\varpi\bigl(\tilde{X}(t)\bigr)\Bigr] \\
& \quad +\dfrac{1}{\gamma_1}Z^{\mathrm{T}}(t)\Bigl[\bigl(G_1^+A_1^*G_1\bigr)^{\mathrm{T}}(\mathscr{K}_M \tilde{z})(t,v^\star) +\gamma_1 G_1^{\mathrm{T}}\tilde{X}(t)\Bigr] \\
& \quad -\dfrac{1}{\gamma_1}Z^{\mathrm{T}}\bigl(G_1^+A_1^*G_1\bigr)^{\mathrm{T}}(\mathscr{K}_M M)(v^\star)\varpi\bigl(\tilde{X}(t)\bigr) ,  \quad t \in (0,T).  
\end{split}
\end{align}
The first cross term appearing in \eqref{eq:finalDer} may be compensated for exactly as done in Section \ref{sect:Step3}, producing
\begin{align}\label{eq:finalDer2}
\begin{split}
 & \dot{V}(t)  \leq -\gamma_0\rho V_0(t) -\gamma_2\biggl(V_1(t) +\dfrac{1}{\gamma_1(v^\star)}V_2(t)\biggr) \\
& \quad +  \dfrac{\theta}{\gamma_1}\int_0^1 \zeta^{\mathrm{T}}(\xi,t)\mathscr{Q}(\xi)\Bigl(\Sigma\bigl(v(X(t),U(t))\bigr)-\Sigma(v^\star)\Bigr)\\
& \quad \times  \bigl[z^\star(\xi) + (\mathscr{K}_2z^\star)\bigr]\dif \xi \\
& \quad +\dfrac{\theta}{\gamma_1} \int_0^1 \zeta^{\mathrm{T}}(\xi,t)\mathscr{Q}(\xi)\Bigl(\Sigma\bigl(v(X(t),U(t))\bigr)-\Sigma(v^\star)\Bigr) \\
& \quad \times \bigl[M(\xi,v^\star) + (\mathscr{K}_2M)(v^\star)\bigr]\varpi\bigl(X_\delta(t)\bigr)\dif \xi \\
& \quad +\dfrac{1}{\gamma_1} Z^{\mathrm{T}}(t)\Bigl[A_2\tilde{X}(t)-\Psi^{-1}(v^\star)\varpi\bigl(\tilde{X}(t)\bigr)\Bigr] \\
& \quad +\dfrac{1}{\gamma_1}Z^{\mathrm{T}}(t)\Bigl[\bigl(G_1^+A_1^*G_1\bigr)^{\mathrm{T}}(\mathscr{K}_M \tilde{z})(t,v^\star) +\gamma_1 G_1^{\mathrm{T}}\tilde{X}(t)\Bigr] \\
& \quad -\dfrac{1}{\gamma_1}Z^{\mathrm{T}}\bigl(G_1^+A_1^*G_1\bigr)^{\mathrm{T}}(\mathscr{K}_M M)(v^\star)\varpi\bigl(\tilde{X}(t)\bigr) ,  \quad t \in (0,T),  
\end{split}
\end{align}
with $\gamma_2 \triangleq  \min\{q, \frac{\omega}{ \norm{\lambda\ped{max}(\mathscr{Q}(\cdot))}_\infty}\}$.
Moreover, the last three terms may be bounded by noting that there exists $\bar{\eta}_0(v^\star) \in \mathbb{R}_{\geq 0}$ such that
\begin{align}
\begin{split}
& Z^{\mathrm{T}}(t)\Bigl[A_2\tilde{X}(t)-\Psi^{-1}(v^\star)\varpi\bigl(\tilde{X}(t)\bigr)\Bigr] \\
& \quad +Z^{\mathrm{T}}(t)\Bigl[\bigl(G_1^+A_1^*G_1\bigr)^{\mathrm{T}}(\mathscr{K}_M \tilde{z})(t,v^\star) +\gamma_1 G_1^{\mathrm{T}}\tilde{X}(t)\Bigr] \\
& \quad -Z^{\mathrm{T}}\bigl(G_1^+A_1^*G_1\bigr)^{\mathrm{T}}(\mathscr{K}_M M)(v^\star)\varpi\bigl(\tilde{X}(t)\bigr) \\
& \qquad \leq \bar{\eta}_0(v^\star)V_0^{\frac{1}{2}}(t)\sqrt{V_1(t) + \dfrac{1}{\gamma_1(v^\star)}V_2(t)}.
\end{split}
\end{align}
Thus, applying Cauchy-Schwarz and then the generalized Young's inequality for products provides
\begin{align}
\begin{split}
&\dfrac{1}{\gamma_1} Z^{\mathrm{T}}(t)\Bigl[A_2\tilde{X}(t)-\Psi^{-1}(v^\star)\varpi\bigl(\tilde{X}(t)\bigr)\Bigr] \\
& \quad +\dfrac{1}{\gamma_1}Z^{\mathrm{T}}(t)\Bigl[\bigl(G_1^+A_1^*G_1\bigr)^{\mathrm{T}}(\mathscr{K}_M \tilde{z})(t,v^\star) +\gamma_1 G_1^{\mathrm{T}}\tilde{X}(t)\Bigr] \\
& \quad -\dfrac{1}{\gamma_1}Z^{\mathrm{T}}\bigl(G_1^+A_1^*G_1\bigr)^{\mathrm{T}}(\mathscr{K}_M M)(v^\star)\varpi\bigl(\tilde{X}(t)\bigr) \\
& \qquad \leq \dfrac{\varepsilon}{2\gamma_1(v^\star)}V_0(t)  + \dfrac{\bar{\eta}_0^2(v^\star)}{2\gamma_1(v^\star)\varepsilon}\biggl(V_1(t) +\dfrac{1}{\gamma_1(v^\star)}V_2(t)\biggr).
\end{split}
\end{align}
Hence, specifying
\begin{align}
\varepsilon & = \varepsilon(v^\star) \triangleq \dfrac{\bar{\eta}_0^2(v^\star)}{\gamma_1(v^\star)\gamma_2}, \quad \gamma_0 = \gamma_0(v^\star) \triangleq \dfrac{2\varepsilon(v^\star)}{\rho\gamma_1(v^\star)},
\end{align}
gives
\begin{align}\label{eq:finalDer3}
\begin{split}
 & \dot{V}(t)  \leq -\bar{\gamma} V(t)\\
& \quad +  \dfrac{\theta}{\gamma_1}\int_0^1 \zeta^{\mathrm{T}}(\xi,t)\mathscr{Q}(\xi)\Bigl(\Sigma\bigl(v(X(t),U(t))\bigr)-\Sigma(v^\star)\Bigr)\\
& \quad \times  \bigl[z^\star(\xi) + (\mathscr{K}_2z^\star)\bigr]\dif \xi \\
& \quad +\dfrac{\theta}{\gamma_1} \int_0^1 \zeta^{\mathrm{T}}(\xi,t)\mathscr{Q}(\xi)\Bigl(\Sigma\bigl(v(X(t),U(t))\bigr)-\Sigma(v^\star)\Bigr) \\
& \quad \times \bigl[M(\xi,v^\star) + (\mathscr{K}_2M)(v^\star)\bigr]\varpi\bigl(X_\delta(t)\bigr)\dif \xi ,  \quad t \in (0,T),  
\end{split}
\end{align}
with $\mathbb{R}_{>0} \ni \bar{\gamma} \triangleq \frac{1}{2}\min\{\rho,\gamma_2\}$. Finally, recalling the control law \eqref{eq:U_deltaOF}, the last cross terms figuring in \eqref{eq:finalDer3} may be formally bounded exactly as done in Section \ref{sect:Step3}, permitting to infer the existence of $\bar{\eta}_1(v^\star), \bar{\eta}_2(v^\star) \in \mathbb{R}_{\geq 0 }$ such that 
\begin{align}\label{eq:VfinDerDynSF}
\dot{V}(t) \leq - \bigl(\bar{\gamma}-\theta\bar{\eta}_1(v^\star)\bigr) V(t) + \theta\bar{\eta}_2(v^\star)V^{\frac{3}{2}}(t), \quad t \in(0,T).
\end{align}
For what follows, it is beneficial to introduce the Hilbert space $\mathcal{X}^2 \triangleq \mathcal{X}\times \mathcal{X}$ equipped with norm $\norm{(\psi_1(\cdot), \psi_2(\cdot))}_{\mathcal{X}^2}^2\triangleq \norm{\psi_1(\cdot)}_\mathcal{X}^2 + \norm{\psi_2(\cdot)}_\mathcal{X}^2$. 
Theorem \ref{theorem:final2} asserts the corresponding output-feedback result of Theorem \ref{theorem:final}. 
\begin{theorem}\label{theorem:final2}
Consider the ODE-PDE interconnection \eqref{eq:originalSystems}-\eqref{eq:MatricesOOO} under Assumptions \ref{ass:ODEf} and \ref{ass:ODEf2}, in closed loop with the observer \eqref{eq:originalSystemsObs}, along with the control law $U(t) = U^\star + U_\delta(t)$, with $U_\delta(t)$ as in \eqref{eq:U_deltaOF}, and suppose that the target equilibrium $(X^\star,z^\star) \in \mathbb{R}^{2}\times \mathscr{D}(\mathscr{A})$ corresponding to the input $U^\star \in \mathbb{R}^{2}$ is such that $\mathbb{R}^{2} \ni v^\star = A_2X^\star + G_2U^\star$ verifies $\theta\bar{\eta}_1(v^\star) < \gamma$. Then, for all ICs $(X_{0},z_{0},\tilde{X}_0,\tilde{z}_0) \triangleq (X(0),z(\cdot,0),\tilde{X}(0),\tilde{z}(\cdot,0)) \in \mathcal{X}^2$ such that
\begin{align}\label{eq:V0_ini2}
\begin{split}
V(0) & < \biggl(\dfrac{\bar{\gamma}-\theta\bar{\eta}_1(v^\star)}{\theta\bar{\eta}_2(v^\star)}\biggr)^2,
\end{split}
\end{align}
with $V(X_\delta(t),\zeta(\cdot,t),\tilde{X}(t),\tilde{z}(\cdot,t))$ as in \eqref{eq:LyapunovFinal}, the system \eqref{eq:originalSystems} and \eqref{eq:PDEzetaTot2SF} admits a unique mild solution $(X,z,\tilde{X},\tilde{z}) \in C^0(\mathbb{R}_{\geq 0};\mathcal{X}^2)$ satisfying
\begin{align}\label{eq:boundPErf2}
\begin{split}
& \norm{(X(t)-X^\star,z(\cdot,t)-z^\star(\cdot),\tilde{X}(t),\tilde{z}(\cdot,t))}_{\mathcal{X}^2} \\
& \quad \leq \bar{\beta}(v^\star)\eu^{-\bar{\sigma} t} \norm{(X_0-X^\star,z_0(\cdot)-z^\star(\cdot),\tilde{X}_0,\tilde{z}_0)}_{\mathcal{X}^2}, \\
& \qquad \qquad \qquad \qquad \qquad \qquad \qquad \qquad \quad t\in[0,T],
\end{split}
\end{align}
for some $\bar{\beta}(v^\star), \bar{\sigma} \in \mathbb{R}_{>0}$. 

\begin{proof}
The proof is identical to that of Theorem \ref{theorem:final}, and thus omitted.
\end{proof}
\end{theorem}
\begin{remark}
Exactly as for the state-feedback case, the results asserted by Theorem \ref{theorem:final2} are global for linear systems ($\theta = 0$).
\end{remark}

In the next Section \ref{sect:sim}, the proposed backstepping stabilization strategy is tested in simulation.


\section{Simulation results}\label{sect:sim}
\begin{figure}
\centering
\includegraphics[width=0.8\linewidth]{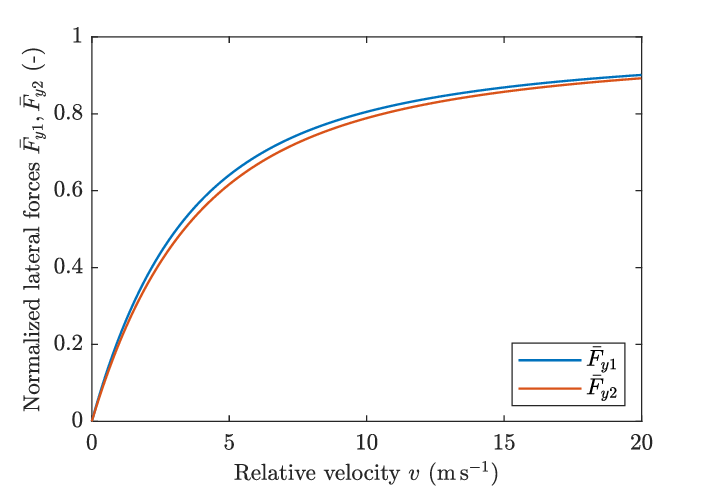} 
\caption{Normalized steady-state lateral tire forces $\bar{F}_{yi} \triangleq \frac{F_{yi}}{2F_{zi}}$, $i \in \{1,2\}$. Model parameters as in Table \ref{tab:parameters}.}
\label{figure:normForc}
\end{figure}

The numerical values for the model parameters of the example discussed below are similar to those reported in \textcite{Guiggiani,SemilinearV} and are listed in Table \ref{tab:parameters}. With the given combination of parameters, Assumptions \ref{ass:ODEf} and \ref{ass:ODEf2} are all fulfilled. Moreover, the considered vehicle is oversteer, and hence inherently unstable for values of the longitudinal speed beyond a critical value. Using the parameter values reported in Table \ref{tab:parameters}, the trends of the normalized steady-state lateral tire forces $\mathbb{R}_{>0} \ni \bar{F}_{yi} \triangleq \frac{F_{yi}}{2F_{zi}}$, $i \in \{1,2\}$, with respect to the rigid relative velocity are plotted in Figure \ref{figure:normForc}. 

In this context, the following numerical results refer to simulations conducted in MATLAB/Simulink\textsuperscript{\textregistered} environment. The semilinear PDE subsystem was first discretized in space using a finite difference scheme with a discretization step of $0.02$, and the resulting ODEs were then simulated in Simulink\textsuperscript{\textregistered} using \texttt{ode4} (Runge-Kutta) with a fixed time step of $10^{-6}$ s\footnote{Such a small time step was needed to handle the closed-loop implementation in the presence of noise, whereas the open-loop system could also be simulated with a time step of $10^{-3}$ s.}. The ICs for the actual system were set to $X_0 = [1.5\; -0.25]^{\mathrm{T}}$, and $z_0(\xi) = [0.003\; 0.003]^{\mathrm{T}}$ (corresponding to $\norm{z_0(\cdot)}_{L^2((0,1);\mathbb{R}^2)} = 0.0042$), whereas those for the observer to $\hat{X}_0 = [0\; 0]^{\mathrm{T}}$, and $\hat{z}_0(\xi) = [0\; 0]^{\mathrm{T}}$. A control input delay of $\delta_U = 0.2$ s was introduced, and the measurements were corrupted with additive white noise. In particular, the lateral velocity and yaw rate components of the rigid relative velocity measurements were perturbed with white noise having standard deviations of 0.5 $\text{m}\,\text{s}^{-1}$ and 0.1 $\text{rad}\,\text{s}^{-1}$, respectively, and sample times of 0.01 and 0.005 s. The larger value was intentionally exaggerated to test robustness against external disturbances, whilst the smaller value reflects the noise level typically observed in standard automotive yaw rate sensors.
\begin{table}[ht]
\centering
\small
\resizebox{\columnwidth}{!}{%
\begin{tabularx}{\columnwidth}{|c|X|c|c|}
\hline
\textbf{Parameter} & \textbf{Description} & \textbf{Unit} & \textbf{Value} \\
\hline 
$v_x$ & Longitudinal speed & $\textnormal{m}\,\textnormal{s}^{-1}$ & $50$ \\ 
$m$ & Vehicle mass & kg & 1300 \\ 
$I_z$ & Vertical moment of inertia  & $\textnormal{kg}\,\textnormal{m}^{2}$ & 2000 \\
$l_1$ & Front axle length & m & 1.4  \\
$l_2$ & Rear axle length & m & 1 \\
$F_{z1}$ & Front vertical force & N & $2.66 \cdot 10^3$ \\
$F_{z2}$ & Rear vertical force & N & $3.72 \cdot 10^3$ \\
$F\ped{w}$ & Lateral wind force & N & $-500$ \\
$l\ped{w}$ & Wind force offset & m & $-0.3$ \\
$L_1$ & Front contact patch length & m & 0.11 \\
$L_2$ & Rear contact patch length & m & 0.09 \\
$\sigma_{1}$ & Front micro-stiffness & $\textnormal{m}^{-1}$ & 240 \\
$\sigma_{2}$ & Rear micro-stiffness & $\textnormal{m}^{-1}$ & 269 \\
$\phi_{1}$ & Front structural parameter & - & 0.92 \\
$\phi_{2}$ & Rear structural parameter & - & 0.92 \\
$\mu_1(\cdot)$ & Front friction coefficient & -& 1\\
$\mu_2(\cdot)$ & Rear friction coefficient & -& 1\\
$a_1$ & Front pressure parameter & - & 0.1 \\
$a_2$ & Rear pressure parameter & - & 0.1 \\
$\theta$ & Friction parameter & - & 1 \\
$\varepsilon$ & Regularization parameter & - & 0 \\
\hline
\end{tabularx}
}
\caption{Model parameters}
\label{tab:parameters}
\end{table}

\begin{figure}
\centering
\includegraphics[width=0.8\linewidth]{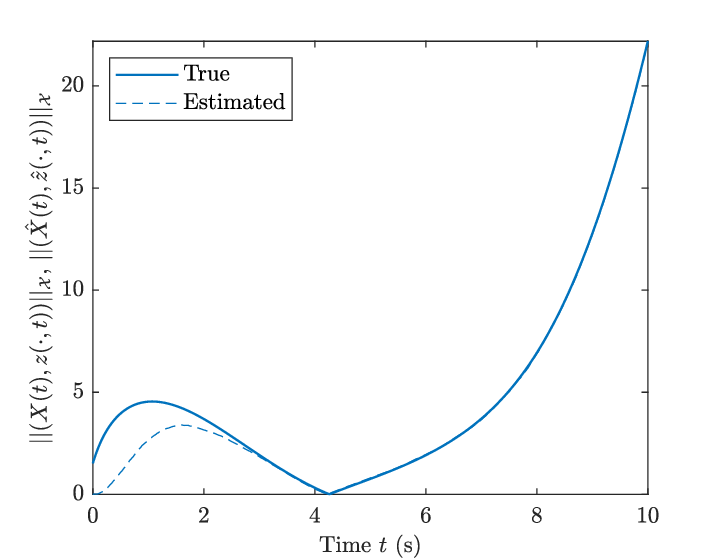} 
\caption{True states (solid tick blue line) and observer estimates (dashed blue line) for the open loop system \eqref{eq:originalSystems}-\eqref{eq:MatricesOOO}, with $p=2$.}
\label{figure:error}
\end{figure}

\begin{figure}
\centering
\includegraphics[width=0.8\linewidth]{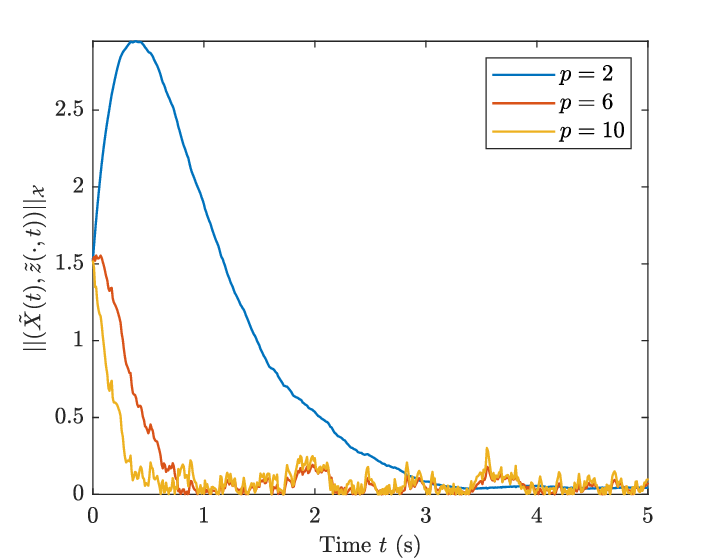} 
\caption{Open-loop convergence of the observer error estimate for three different values of observer gains $p=2$, 6, and 10.}
\label{figure:parameters2}
\end{figure}

Figure \ref{figure:error} illustrates the unstable behavior of the uncontrolled vehicle driving at $v_x = 50$ $\textnormal{m}\,\textnormal{s}^{-1}$. The observer synthesized as in Section \ref{sect:OutFedd}, with $L_1 \in \mathbf{M}_2(\mathbb{R})$ in \eqref{eq:originalSystemsODEObs} specified as $L_1 = -(A_1 + pI_2)A_2^{-1}$ ($p=2$ in Figure \ref{figure:error}), predicts the true states with great accuracy, with the estimates converging approximately for $t = 3$ s. 
In particular, concerning the open loop dynamics, the effect of the observer gain on the convergence rate of the observer errors $\tilde{X}(t)$ and $\tilde{z}(\xi,t)$ is shown in Figure \ref{figure:parameters2} for $p = 2$, $6$, and $10$. As it might be intuitively expected, inspection of Figure \ref{figure:parameters2} reveals that higher values of $p$ produce faster convergence rates, but exert a less satisfactory filtering action.

\begin{figure}
\centering
\includegraphics[width=0.8\linewidth]{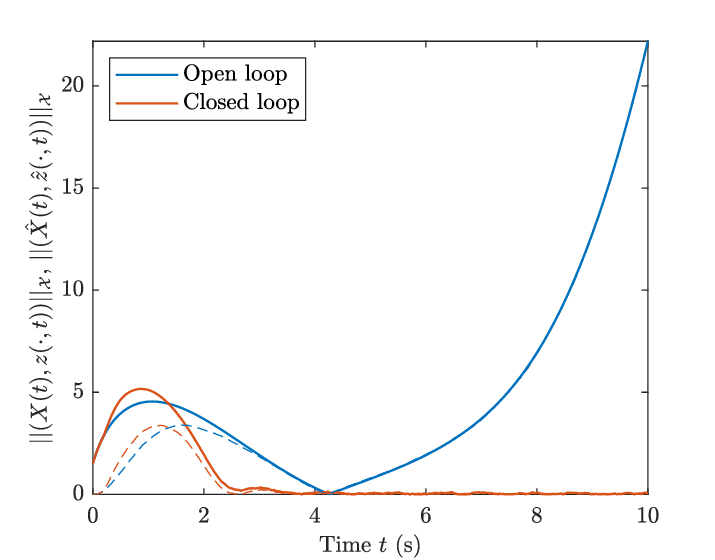} 
\caption{Open and closed loop behaviors of the ODE-PDE system described by \eqref{eq:rigid}-\eqref{eq:PressureDistr}, with $p=q = 2$.}
\label{figure:parameters3}
\end{figure}

\begin{figure}
\centering
\includegraphics[width=0.8\linewidth]{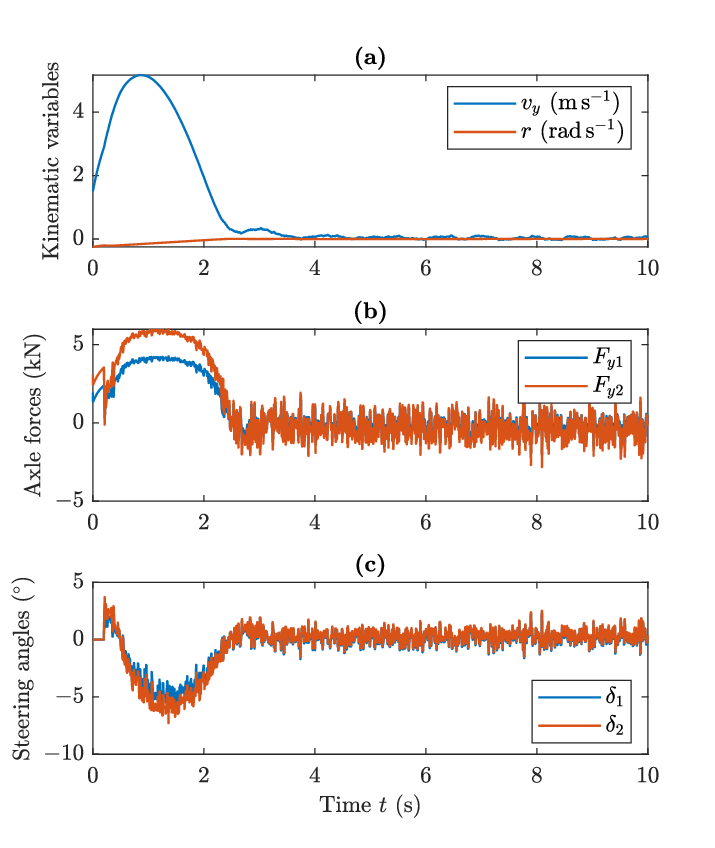} 
\caption{Closed loop behavior of the lumped states and steering inputs, for $p=2=q = 2$: \textbf{(a)} kinematic variables; \textbf{(b)} axle forces; \textbf{(c)} steering inputs.}
\label{figure:lumpedStates}
\end{figure}

The open and closed loop behaviors of the ODE-PDE system described by \eqref{eq:originalSystems}-\eqref{eq:MatricesOOO} are compared in Figure \ref{figure:parameters3} for $p=2=q = 2$. Despite an initial overshoot, the output-feedback controller designed as in Section \ref{sect:OutFedd} successfully stabilizes the vehicle around the desired equilibrium  $(X^\star, z^\star) = (0, z^\star)$, with $z^\star(\xi)$ solving \eqref{eq:equilibria}, at $t \approx 2$ s, preventing the norm $\norm{(X(t),z(\cdot,t))}_{\mathcal{X}}$ from exceeding a maximum threshold of 5.
The closed loop trends of the kinematic variables, axle forces, and steering inputs are depicted in Figure \ref{figure:lumpedStates}\textbf{(a)}, \textbf{(b)}, and \textbf{(c)}. It may be observed that both the states $v_x(t)$ and $r(t)$ nearly converge to zero, whereas, after an initial transient extinguished at $t \approx 2$ s, the tire forces are characterized by a noisy dynamics around the the steady-state values $[F_{y1}^\star\; F_{y2}^\star]^{\mathrm{T}} = -[146\; 354]^{\mathrm{T}}$, which correspond to the stationary input $U^\star = [\delta_1^\star\; \delta_2^\star]^{\mathrm{T}} = [0.12\; 0.24]^{\mathrm{T}}$ needed to compensate for the action of the wind gust $F\ped{w}$. Similarly, the steering angles exhibit a pronounced transient, but never exceed $5^\circ$ in absolute value, confirming \emph{a posteriori} the feasibility of the maneuver. 

\begin{figure}
\centering
\includegraphics[width=0.8\linewidth]{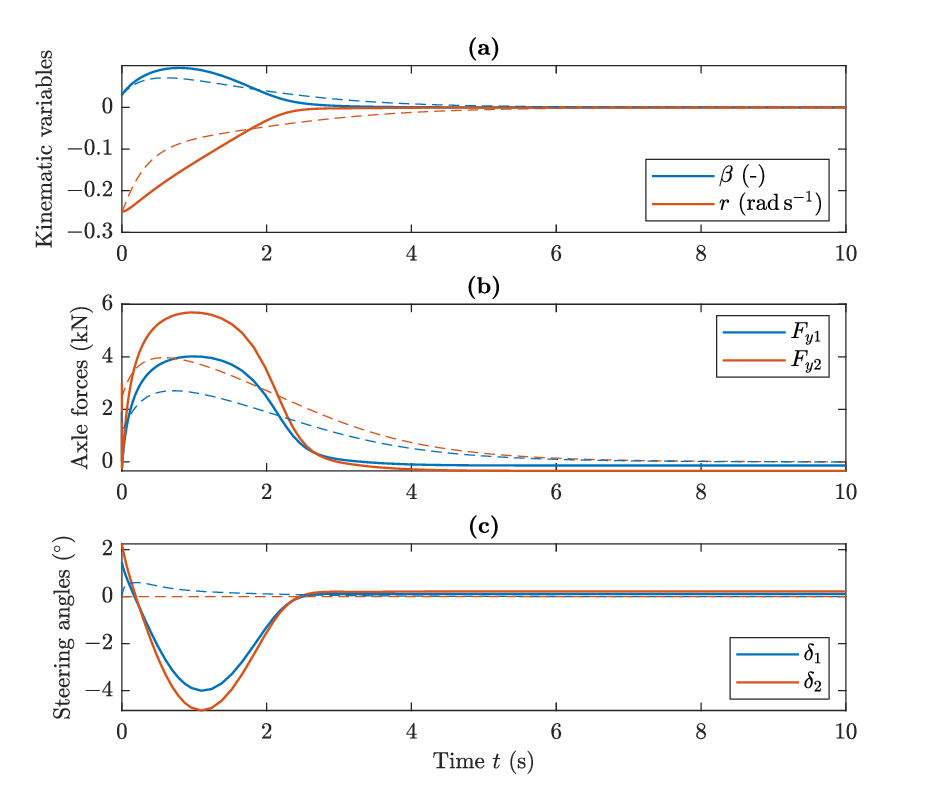} 
\caption{Comparison between the proposed infinite-dimensional controller (solid thick lines) and a simpler implementation based on a reduced-order model (dashed lines) for $v_x = 50$ $\mathrm{m}\,\mathrm{s}^{-1}$.}
\label{figure:comparisonROM}
\end{figure}

Figure \ref{figure:comparisonROM} compares the performance of the controller synthesized according to the methodology proposed in this paper with that of a simpler implementation based on a reduced-order model obtained by neglecting the PDE dynamics. In Figure \ref{figure:comparisonROM}, the quantity $\mathbb{R} \ni \beta(t)\triangleq v_y(t)/v_x$ denotes the vehicle \emph{sideslip angle}. The gains of the reduced-order controller were tuned to achieve performance comparable to that of the infinite-dimensional one (for clarity, measurement noise and delay effects were neglected). One advantage of the reduced-order approach is that vehicle stabilization can be achieved using front steering alone, whereas the infinite-dimensional implementation requires actuation at both the front and rear wheels. It is also interesting to note that, during the initial transient, the infinite-dimensional controller commands a steering action in the opposite direction to that generated by the reduced-order controller.

\begin{figure*}
\centering
\includegraphics[width=0.7\linewidth]{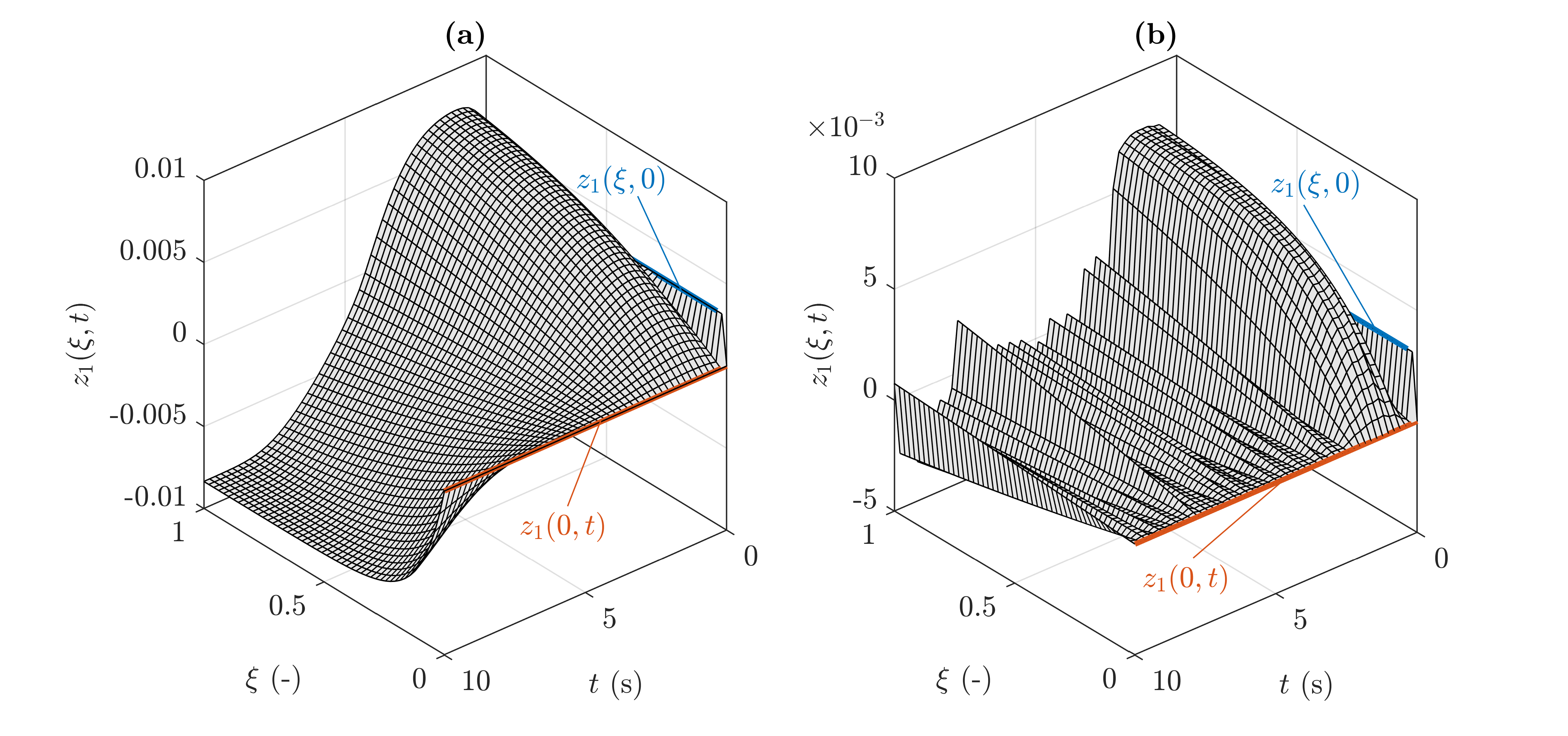} 
\caption{Evolution of the PDE state $z_1(\xi,t)$, along with its IC (blue lines) and BC (orange lines): \textbf{(a)} the open loop case; \textbf{(b)} closed loop case with $p=2=q = 2$.}
\label{figure:PDEs}
\end{figure*}

Similar considerations as those reported above may be drawn by inspecting Figure \ref{figure:PDEs}, where the open and closed loop dynamics of the PDE state $z_1(\xi,t)$ are juxtaposed (Figure \ref{figure:PDEs}\textbf{(a)} and \textbf{(b)}, respectively). Specifically, it may be observed that, under the action of the control input \eqref{eq:U_deltaOF}, the distributed state rapidly approaches its steady-state profile around $t = 2$ s, which is consistent with the observations reported above. Also in this case, residual oscillations should be ascribed to the noisy measurements, which affect the quality of the estimate $\hat{z}_1(\xi,t)$. Additional simulations were conducted by varying the gains $p$ and $q$, without appreciable differences in the qualitative behavior of the closed loop system. In practice, however, the gain $q$ cannot be chosen arbitrarily large, due to the physical and mechanical constraints on the admissible steering angles. 

\begin{figure}
\centering
\includegraphics[width=0.8\linewidth]{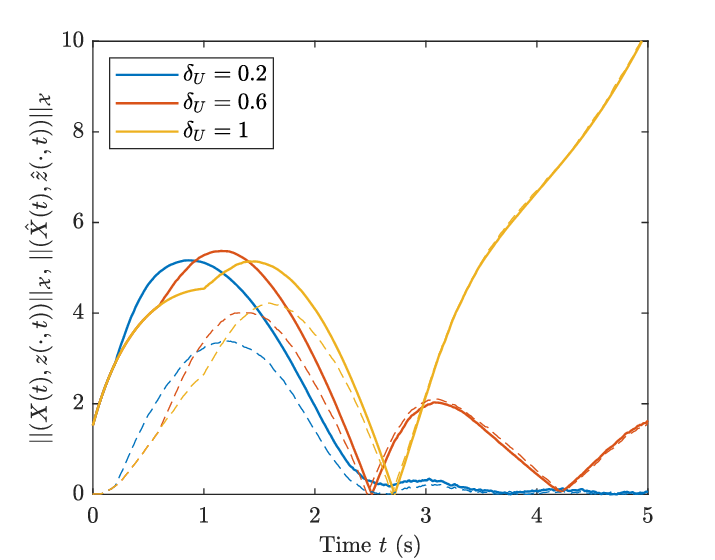} 
\caption{Closed loop behavior of $\norm{(X(t),z(\cdot,t))}_{\mathcal{X}}$ and $\norm{(\hat{X}(t),\hat{z}(\cdot,t))}_{\mathcal{X}}$ (solid and dashed lines, respectively) for different values of the input delay $\delta_U = 0.2$, 0.6, and 1.}
\label{figure:delay}
\end{figure}

\begin{figure}
\centering
\includegraphics[width=0.8\linewidth]{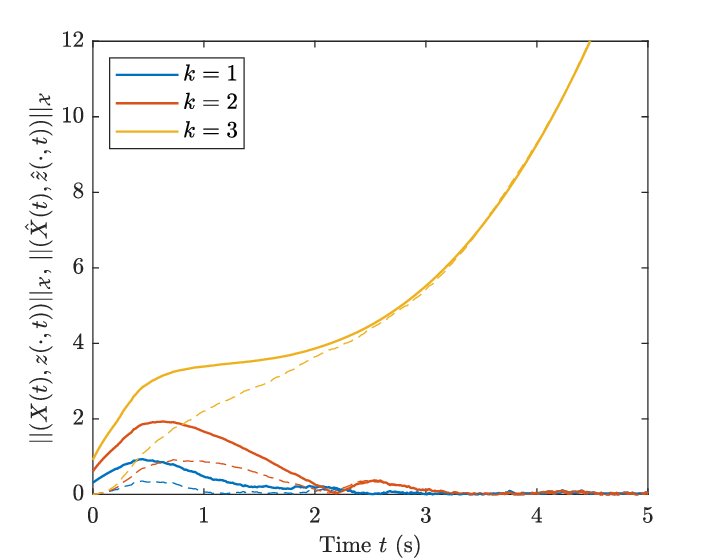} 
\caption{Closed loop behavior of $\norm{(X(t),z(\cdot,t))}_{\mathcal{X}}$ and $\norm{(\hat{X}(t),\hat{z}(\cdot,t))}_{\mathcal{X}}$ (solid and dashed lines, respectively) for different values of $k = 1$, 2, and 3.}
\label{figure:ICs}
\end{figure}

To further test the robustness of the proposed controller, the effect of different time delays $\delta_U$ and ICs $X_0$ was also investigated. Figure \ref{figure:delay} depicts the dynamics of the norms $\norm{(X(t),z(\cdot,t))}_{\mathcal{X}}$ and $\norm{(\hat{X}(t),\hat{z}(\cdot,t))}_{\mathcal{X}}$ (solid and dashed lines, respectively), for different values of the input delay $\delta_U = 0.2$, 0.6, and 1 s. In particular, it may be observed that, whilst an input delay of 0.6 s still manages to stabilize the system's dynamics (albeit producing large undesired oscillations), a delay of 1 s causes an unstable closed loop response. The influence of different initial conditions $X_0 = -k[0.3\; -0.05]^{\mathrm{T}}$ on the closed loop dynamics of the ODE-PDE system is instead illustrated in Figure \ref{figure:ICs} for $k=1$, 2, and 3. As expected from Theorems \ref{theorem:final} and \ref{theorem:final2}, ICs that are further from the target equilibrium may invalidate the effectiveness of the proposed stabilization strategy and introduce dangerous instabilities, as it happens for $k=3$.

\section{Conclusions}\label{sect:conclusion}
This paper presented a passivity-exploiting backstepping stabilization method for semilinear single-track vehicle models with distributed tire friction dynamics. By leveraging the strict dissipativity of the PDE subsystem, a Lyapunov functional was constructed, ensuring local exponential stability under mild structural assumptions. The proposed methodology accommodates both state-feedback and output-feedback designs, the latter being enabled by a cascaded observer whose convergence was proven using a separable Lyapunov functional.

The approach was validated numerically considering non-ideal scenarios accounting for external disturbances and uncertainties. Simulation results demonstrated that the controller effectively achieves stabilization to the desired equilibrium for an oversteer vehicle driving above the critical speed, highlighting its relevance for automotive safety and performance. In this context, it is worth mentioning that, whilst the obtained stability conditions are locally valid and somewhat conservative, they provide a rigorous foundation for future developments. Extensions of this work may focus on reducing conservativeness and establishing global stability guarantees. Moreover, whereas this paper was restricted to single-track models, more refined representations with additional degrees of freedom, such as roll, pitch, and heave, might be considered. In this case, alternative control implementations exploiting the time-scale separation between the vehicle's rigid body motion and tire dynamics could be developed. Combined braking and steering actuation strategies may also be investigated to overcome the all-wheel-steering assumption formulated in this work and address more realistic maneuvers. Finally, further research could also explore the integration of the proposed controllers with advanced estimation schemes and its application to other rolling contact systems.


\subsubsection*{Acknowledgments}
This research was financially supported by the project FASTEST (Reg. no. 2023-06511), funded by the Swedish Research Council.

\subsubsection*{Declaration of interest}
Declaration of interest: none.

{\setstretch{1} 
\printbibliography
}

\end{document}